\documentstyle[aps,prb,epsfig]{revtex}

\begin{document}

\draft


\title{Itinerant Electron Ferromagnetism in the Quantum Hall Regime}
\author{Marcus Kasner$^{1,2}$, J.J. Palacios$^{1,3}$ and A.~H.~MacDonald$^{1}$}
\address{$^{1}$Dept.\ of Physics,
        Indiana University, Swain Hall West 117, 
Bloomington, IN 47405, USA \\
$^{2}$Institut f\"ur Theoretische Physik, Otto-von-Guericke Universit\"at, PF 4120, 
D-39016 Magdeburg, Germany \\
$^{3}$Department of Physics and Astronomy,
University of Kentucky, Lexington KY 40506-0055, USA
}


\maketitle

\begin{abstract}
We report on a study of the temperature and Zeeman-coupling-strength
dependence of the one-particle Green's function of a two-dimensional  
(2D) electron gas at Landau level filling factor $\nu =1$ where 
the ground state is a strong ferromagnet.
Our work places emphasis on the role played by 
the itinerancy of the electrons, which carry the spin magnetization
and on analogies between this system and conventional 
itinerant electron ferromagnets.  We discuss the application
to this system of the self-consistent Hartree-Fock approximation,
which is analogous to the band theory description of metallic ferromagnetism 
and fails badly at finite temperatures because it does not 
account for spin-wave excitations.  We go beyond this level 
by evaluating the one-particle Green's function 
using a self-energy, which accounts for quasiparticle spin-wave
interactions.  We report results for the temperature dependence of the 
spin magnetization, the nuclear spin relaxation rate, and
2D-2D tunneling conductances.  Our calculations predict a sharp peak
in the tunneling conductance at large bias voltages 
with strength proportional to temperature.
We compare with experiment, where available, and with 
predictions based on numerical exact diagonalization
and other theoretical approaches.

\end{abstract}


\widetext

\section{Introduction}

The physics of a two-dimensional electron system (2DES) 
in a magnetic field is in many respects unique. 
Since the degeneracy of the discrete Landau levels 
increases in proportion to the magnetic field strength, 
all electrons can be accommodated in the lowest Landau level (LLL) for
sufficiently strong fields.
A Landau level then behaves, in many respects, like a band 
of zero width and the system can be regarded 
as the extreme limit of a strongly correlated narrow-band
electronic system.  In this paper we focus on the 
spin magnetism of the electron system at Landau level 
filling factor $\nu = 1$ corresponding, when the spin degree
of freedom is accounted for, to the case of a half-filled `Landau band'. 
At this filling factor, it is known\cite{MFB96,GM96}
that, when disorder can
be neglected, the ground state of the 2DES is ferromagnetic. 
We are motivated to study this system because it shares many  
of the difficulties\cite{moriya}, which have confounded attempts to
build a complete theory of metallic ferromagnetic systems, yet is free of the 
troubling but incidental consequences of a complex band structure.  
The finite temperature properties
of metallic ferromagnets are more involved than those of
insulating ferromagnets because of the importance of both 
spin and charge degrees of freedom, so much so that 
much early theory was based on misguided attempts to 
assign magnetic and conducting properties to separate 
classes of electrons.  Despite an immense effort and
many advances\cite{moriya},
no completely satisfactory theory of metallic ferromagnets exists.
Progress in understanding metallic ferromagnets has been
hampered in part by the quantitative importance of details
of the electronic band-structure, which may not be accurately known
or may be difficult to render faithfully in going beyond mean-field
theories of many-body effects.
The present system has no such difficulties.

Our work is also motivated by recent experimental progress.
Barrett {\it et al.} \cite{BDPWT95}  
were able to measure the temperature-dependence of the spin-polarization 
and nuclear-spin relaxation rates at 
fixed filling factors around $\nu = 1$ using nuclear magnetic
resonance (NMR) techniques.  Later, Manfra {\it et al.} \cite{MAGBPW96}
extracted the spin magnetization from magneto-optical absorption measurements. 
It is our hope that critical comparison between experiment
and theory will yield insights with wider relevance
to the finite temperature properties of itinerant electron ferromagnets. 
It is, however, important to recognize that the
2DES at $\nu =1$ is different from conventional itinerant electron ferromagnets
in several important ways.  Most importantly, two-dimensionality implies that
its spontaneous spin magnetic moment will not survive at finite temperatures 
($T_c = 0$.)  In addition, the strong magnetic field, which through its
coupling to the electron's orbital degrees of freedom produces Landau levels,
also produces a Zeeman coupling to electron's spin degree of freedom.
For the most studied 2DES's, those formed at GaAs/AlGaAs heterojunctions,
the Zeeman coupling is quite small compared to both Landau level separations
and the characteristic interaction energy scale.  As we discuss
below, the main effect of Zeeman coupling at low temperatures is 
to cut-off the decrease of the magnetization due to the thermal excitation of
very long wavelength spin-waves and mitigate consequences of
the system's reduced dimensionality.

The recent experimental work has stimulated 
two different theoretical approaches, which focus on 
different aspects of the spin-magnetization physics.  
Read and Sachdev\cite{RS95} have compared experimental data  
with large $N$ limits of a quantum continuum field theory model, 
which provides an accurate description of the long-wavelength
collective behavior of the electronic spins. 
In this theory, physical properties are dependent only
on the two independent ratios between the thermal energy, $k_B
T$, the Zeeman coupling strength, $\Delta_z$, and 
the spin-stiffness energy, $\rho_s$.  Recently this work
has been extended by Timm {\it et al.}.\cite{timm}  The field-theory 
description is expected\cite{remarkrhos} to be accurate at low temperature 
when the Zeeman coupling strength is weak.
This approach achieves reasonable overall agreement between
theory and experiment, at least at low temperatures.   
Our work has somewhat different motivation and follows
a different line.  We are interested in addressing the 
temperature dependence of the underlying electronic structure,
as it changes in concert with the change in the spin-magnetization.
Hence we focus on the one-particle Green's function.
A brief account of some parts of this work has been
published previously.\cite{KM96} From the Green's function 
we can calculate the electronic spectral function and 
hence the magnetization, the tunneling density of states,
and (if vertex corrections are neglected) the nuclear-spin
relaxation time.  The approximation we use is one, which accounts for 
the interaction of quasiparticles with thermally excited spin-waves.
This approximation has deficiencies.
At low-temperatures and at low Zeeman energies the magnetizations
we calculate do not appear to be in quite as good agreement with experiment
as magnetizations from large $N$ approximations in the 
field-theory calculations.  At medium and high temperatures, we do not account 
systematically for temperature-dependent screening effects which
are likely to be important.  Some progress in the latter direction has recently
been reported by Haussmann\cite{Hau96},  whose bosonized self-consistent 
random-phase approximation yields satisfying
results at high temperatures, but fails at low temperatures. 
Progress on these fronts, which can be checked by comparison 
with experiment, may suggest routes toward more generically satisfactory
theories of itinerant electron ferromagnets. 

Our paper is organized as follows.  In Section II we briefly review 
established results for the ground
state and elementary excitations of the 2DES at $\nu =1$, which 
will be important for subsequent discussion.  The ground
state has all spins aligned by an arbitrarily weak Zeeman coupling. 
If we neglect Landau level mixing, and we do throughout this paper,
this state has no pure charge excitations.
Its elementary excitations all have a single reversed spin.
It turns out that, in the quantum Hall regime,
the Hartree-Fock approximation (HFA) 
is exact for the ground state and the time-dependent Hartree-Fock
approximation is exact for its elementary excitations.
The situation is therefore similar to that for many typical metallic
ferromagnets, where there is substantial evidence that the
ground state is well described by the Hartree-Fock-like
Kohn-Sham equations\cite{JG89} of the spin-density-functional formalism
and that its elementary excitations are well-described by 
the time-dependent generalization of density-functional theory.
In Section III we discuss the application of 
the self-consistent Hartree-Fock approximation (SHF) 
at finite temperature. The failure of this approximation at finite 
temperatures is analogous to the well known failure of 
the band theory of magnetism to provide even a rough account  
of the ordering temperature.  The approximation, on which the 
present work is based, is discussed in 
Section IV.  We obtain an expression for the electronic 
self-energy by analytically evaluating a particle-hole ladder
summation involving Green's functions of opposite spin.
We emphasize that in our microscopic theory, it is essential to account
for screening even in the low-temperature limit.
Section V discusses results for the spin magnetization,
the spin-lattice relaxation rate, and the temperature dependent 
2D-2D tunneling $I-V$ relation, all based on this self-energy approximation. 
In this section we also compare our results with
available experiments, with other approximate theories, and 
with data from finite size numerical calculations of the magnetization
and the magnetic susceptibility.  
We conclude that it is necessary to account for the finite
thickness of the quantum well in comparing with experiment
and that the magnetization will be overestimated at 
high temperatures by models, which do not account for 
electronic itinerancy.   We predict the occurrence of a sharp peak, 
with strength approximately proportional to temperature, 
in the tunneling $I-V$ relation at when $eV$ is close
to the zero-temperature spin-splitting. 
In Section VI we discuss some aspects of our calculation which 
point to difficulties in developing a completely satisfactory 
microscopic theory.  Finally, we conclude in Section VII with 
a brief summary.

\section{Strong Field Limit Preliminaries}

In this paper we deal with a two-dimensional electron system where 
the electrons are subject to a constant perpendicular magnetic field
of strength $\vec{B}= B \vec{e_{z}}$. We are particularly interested 
in the situation at filling factor $\nu=1$ so that the number of electrons 
in the system ($N$) equals the number of single-particle
orbitals available in each Landau level; $N_{\phi}  = A B / \Phi_0$. 
$A$ is the area of the two-dimensional
system and $\Phi_0=h/e$ is the magnetic flux quantum. 
One of the underlying assumptions in our model is 
that the Landau level separation is 
large enough that we can ignore fluctuations in which electrons occupy 
higher orbital Landau levels.
This requires that the Zeeman gap and the interaction strength 
be sufficiently small in comparison to $\hbar \omega_c$ where 
$\omega_c = e B/ m^*$ is the cyclotron frequency. 
The microscopic Hamiltonian in the Landau gauge ($\vec{A} = (0,Bx,0)$) 
is then\cite{HF84} 
\newpage 
\begin{eqnarray}
 H & = & - \frac{1}{2} \Delta_{z}(N_{\uparrow}-N_{\downarrow})
 \nonumber \\ & +  & \frac{\lambda}{2}
 \sum\limits_{p,p^{'},q \atop \sigma, \sigma^{'}}
 \tilde{W}(q,p-p^{'}) c^{\dagger}_{p+\frac{q}{2},\sigma}
 c^{\dagger}_{p^{'}-\frac{q}{2}, \sigma^{'}} c_{p^{'}+\frac{q}{2}, \sigma^{'}}
 c_{p-\frac{q}{2}, \sigma}
\label{H}
\end{eqnarray}
where $\lambda= e^{2}/(4\pi \epsilon \ell_{c})$ is the interaction 
coupling constant, $\Delta_z = | g \mu_{B}B|$ is the Zeeman energy, and $p=k_{y}=2 \pi l/L_{y}$, ($l=0,\pm 1, \ldots$) 
are the momenta in $y$-direction.
The magnetic length $\ell_{c}=\sqrt{\hbar/|eB|}$ is used as the unit of
length below.  When Landau-level mixing is neglected the kinetic energy term
in the Hamiltonian is constant ($N \hbar \omega_c /2$) and 
is therefore neglected. 
In Eq.~(\ref{H}), $N_{\sigma}=\sum_{p}c^{\dagger}_{p,\sigma}c_{p,\sigma}$
is the number operator for an electron with spin $\sigma$.
We choose $\sigma = \uparrow$ as the direction parallel to 
the external magnetic field. 

In the Landau--gauge the two--particle matrix element 
is given by\cite{HF84} 
\begin{equation}
 \tilde{W}(q,p-p^{'})= \frac{1}{L_y} \int_{-\infty}^{\infty} \frac{d k_x}{2\pi}
 \tilde{V}(k_x,q) e^{-(\frac{k_x^{2}+q^2}{2})\ell_c^2}
 e^{i k_{x}(p-p^{'})\ell_c^2} 
\end{equation}
where $\tilde{V}({\vec k})$ is the Fourier transform
of the effective 2D Coulomb interaction, which may include
modifications to account for the finite thickness of 
the quantum well containing the electrons or, as we discuss
below, the {\it ad hoc} incorporation of screening effects. 
The interaction vertex, shown graphically in Fig.~\ref{vert},
is a function only of $k_y=q_3-q_2=q$, the momentum transfer 
due to the interaction, and $q_1-q_3=p-p'$,
the momentum difference of the interacting particles. 
In the physically realistic case of long-range Coulomb interactions
between the electrons it is necessary to incorporate a neutralizing 
positive background by setting $\tilde{V}(k_x,k_y) \propto 
\sum_p \tilde{W}(q =0, p) =0$.  
Note that the interaction term is similar to that of a
one-dimensional interacting fermion model with spin-split bands
of zero width. However, in the present case 
$\tilde{W}$ depends not only on the transferred
momentum $q$, but also on the momentum difference of the 
incoming or outgoing particles. 
This would be true even if we choose a hard core interaction in
real space $V(\vec{r})=4\pi V_{0}\delta^{(2)}(\vec{r}/\ell_c)$.
For this model, the matrix element $\tilde{W}(q,p-p^{'})$
becomes $V_{0}\sqrt{2/\pi}e^{-(q^2/2-(p-p')^2)\ell_c^2/2}$. 

Unlike most interacting electron systems, screening of mutual 
interactions does not play a major role in the correlation
physics of a quantum Hall ferromagnet, at least at low temperatures.
In fact, static screening in this limit is weak because of 
the gap for charged excitations of the ground state.
Nevertheless, we obtain below the somewhat surprising result that 
the electron self-energy has a weak divergence if screening is 
completely neglected.  For that reason 
we allow for the {\em ad hoc} inclusion of screening 
effects in our calculations 
by substituting for the Fourier transform of the Coulomb interaction, 
$\tilde{V}_{c}(\vec{k}) = 2\pi \ell_c/k $, the local static screening 
form $\tilde{V}(\vec{k})= 2\pi \ell_c/(k+k_{sc})$.
The constant value assigned to the screening wave vector
$k_{sc}$ is discussed below.  

The Green's functions we calculate at a particular temperature depend,
up to an overall energy scale, 
only on the ratio $\Delta_z/\lambda$.  At typical field strengths 
this ratio is small, even though  
$\lambda$ varies approximately as $B^{1/2}$ while $\Delta_z$ is 
proportional to $B$.  For example\cite{BDPWT95} at a magnetic field
of $B=7 T$, $\Delta_z/\lambda \simeq 2.2K/136K = 0.016$.
Thus we will be interested primarily in the 
case where the interaction term dominates over
the one--particle spin-dependent term. 

We now briefly recount some known results for the ground 
state and low--lying excitations of the strong field Hamiltonian.  
We start by considering the case of vanishing Zeeman coupling. 
If the interaction is of the hard-core type, the ground state
at filling factor $\nu=1$ can be determined\cite{MFB96} by identifying 
zero energy eigenstates of the positive definite Hamiltonian.
It turns out that these must be the product of the 
the Slater determinant ($|\Psi_S\rangle$) 
constructed from all $N_{\phi}$ one-particle 
orbitals in the lowest Landau level and a many-particle spinor.
The antisymmetry property of the many-fermion wavefunction then
requires that the spinor be completely symmetric.  From this it follows
that it has total spin quantum number $S = N/2$.   
By this argument we are able to establish with some rigor 
that the ground state at $\nu = 1$ is a strong ferromagnet,
in agreement\cite{remarkhf} with the Hartree-Fock approximation 
discussed in the following section and, if we 
regard the degenerate Landau level as analogous to an open
atomic shell, with Hund's rule arguments.
An infinitesimal Zeeman coupling selects from this spin 
multiplet a ground state, in which all spins are aligned;
for a Zeeman field in the $\hat z$ direction 
the ground state has $S^z = S = N/2$
and the non-degenerate ground state wavefunction is given
exactly by the only state in the $\nu = 1$ many-fermion Hilbert space, 
which has these quantum numbers: 
\begin{equation}
|\Psi_{\nu=1}> = |\uparrow \uparrow  \uparrow  \uparrow  \uparrow  \uparrow
\uparrow
        \uparrow  \uparrow  .... \uparrow > |\Psi_{S}> \; .
\end{equation}
Finite size exact diagonalization calculations\cite{polar,GM95,Kas96}
can be used to establish that these conclusions remain valid for the 
realistic case of Coulomb interactions.  

The simplest neutral excitations of the ferromagnetic ground state are 
those with a single reversed spin. 
It turns out that it is possible to analytically solve for 
the wavefunctions and eigenenergies of these excited states\cite{BIE81,KH84}. 
The normalized eigenstates may be labeled by a wavevector $\vec{k}$ and
in second quantized notation are given by
\begin{eqnarray}
|\vec{k}>= \frac{1}{\sqrt{N}} \sum_{q} e^{-iqk_x\ell_c^2} c_{q,\downarrow}^{\dagger}
c_{q+k_y,\uparrow}^{\vphantom{\dagger}}|\Psi_{\nu=1}>
\label{SW}
\end{eqnarray}
The operator relating this state to the ground state is 
proportional to the Fourier transform of the projection of the 
spin-lowering operator onto the lowest Landau level:
$|\vec{k}> = e^{k^2\ell_c^2/4} \bar{S}^{-}(\vec{k})|\Psi_{\nu=1}>/\sqrt{N}$
where $\bar{S}^{-}(\vec{k})$ is
$S^{-}(\vec{k})=S^{x}(\vec{k})-i S^{y}(\vec{k})$ projected onto the LLL. 
These states appear to be similar to the single magnon states
of a localized-spin ferromagnet, but there is an important distinction, 
which is most easily explained by considering a finite size system.
The dimension of the $S^z = N/2-1$ subspace
is $N_{\phi}^2 = N^2$, since there are $N$ possible
states for the minority spin electron and $N$ possible states for the
majority spin hole.  It is possible to show\cite{RM86} that for 
a finite size quantum Hall system, the number of inequivalent values of 
the wavevector $\vec k$ is also $N^2 $.  Since there is
one wavevector for each state and translational invariance decouples states
with different wavevectors, the eigenstates can be constructed by
symmetry arguments alone.  This should be contrasted with the 
case of localized spin Heisenberg ferromagnets for which 
single spin-flip states can also be constructed by symmetry arguments 
alone, but the number of such states is only $N$.  
The much larger number of states in the present problem occurs
because of the possibility in itinerant electron systems of 
changing the orbital occupied by an electron whose spin has been
reversed.

The nature of the single spin-flip excitations of quantum
Hall ferromagnets gradually
changes from having collective spin wave character at long wavelengths to
having single-particle character at larger wavevectors.\cite{KH84,Mac84}.
This property is reflected by the dispersion relation\cite{BIE81,KH84} for
single spin-flip states: 
\begin{eqnarray}
\epsilon_{SW}(\vec{k}) = \Delta_{z} + \lambda  (\tilde{a}(0) - \tilde{a}(\vec{k}))\;,
\label{DISP}
\end{eqnarray}
where the quantity\cite{limit} 
\begin{equation}
\tilde{a}(\vec{k})= \int \frac{d^{2}\vec{q}}{(2\pi)^2} 
\tilde{V}(\vec{q})e^{-\frac{q^2\ell_c^2}{2}} e^{i(\vec{q} \cdot (\hat z \times \vec{k}))\ell_c^2}.
\label{aofk}
\end{equation}
Eq.~(\ref{aofk}) is easy to understand.\cite{KH84}  It represents the 
attractive interaction between a minority spin 
electron and a majority spin hole, quantum mechanically
smeared over their respective cyclotron orbits, and separated in real
space by $\ell_c^2 \hat z \times \vec k$.  The attractive interaction
contributes negatively to the excitation energy. 
This magnetoexcitonic picture
of spin-flip excitations is especially appropriate 
when the electron-hole separation exceeds 
the cyclotron orbit size, {\it i.e.} when 
$\ell_c |\vec k| > 1$.  We use this picture in Section IV to interpret 
our result for the interaction between electrons and 
spin-waves.  The gap for creating infinitely
separated electron-hole pairs, $\Delta_z + \lambda \tilde{a} (0)$, 
is associated with the incompressible\cite{leshouches} property of quantum 
Hall states.  The property $\epsilon_{SW}(\vec {k} \to 0) =
\Delta_z$ is required by spin-rotation invariance of the interaction
Hamiltonian.  For small $k$ the excitations are collective 
in character and 
\begin{equation}
\epsilon_{SW}(\vec {k}) = \Delta_z + 4 \pi \rho_s \ell_c^2 k^2
\label{longwavelengths}
\end{equation}
where $\rho_s = \lim_{k \to 0} \lambda (\tilde{a}(0) - \tilde{a}(\vec k))/(4 \pi \ell_c^2 k^2 )$
is the spin-stiffness parameter, which appears in field theory phenomenologies.
In Fig.~\ref{disp} we plot $\epsilon_{SW}(\vec{k})$ and the above long-wavelength
approximation for the case of a weakly screened Coulomb interaction.

The elementary charged excitations of the $|\Psi_{\nu =1}\rangle$ are 
also known exactly:  
\begin{eqnarray}
|k\rangle_e &=& c_{k,\downarrow}^{\dagger} |\Psi_{\nu =1}\rangle \nonumber \\
|k\rangle_h &=& c_{k,\uparrow} |\Psi_{\nu =1}\rangle.
\label{chgstates}
\end{eqnarray} 
These single Slater determinant states are the maximally 
spin-polarized states with $N = N_{\phi} \pm 1 $ and have energies 
$ E = E_{\nu=1} + \mu + \xi^{HF}_{\downarrow} $ and 
$ E = E_{\nu=1} - \mu - \xi^{HF}_{\downarrow} $, respectively.
Explicit expressions for the Hartree-Fock energies are given in the 
following section.  Recently it has become clear that 
for systems with weak Zeeman energies, charged Skyrmion\cite{skyrmion}
excitations can have lower energies than these states.  However, 
as we discuss below, Skymion states will have little spectral weight
in the one-particle Green's function. 

\section{The Hartree-Fock Approximation and Shortcomings of 
the Band Theory of Itinerant Electron Ferromagnetism}

In anticipation of subsequent sections we discuss the Hartree-Fock
approximation using the lexicon of 
the imaginary time thermodynamic Green's function
technique.\cite{FW71,NO88,Mah90} 
When only the one--particle Zeeman term is retained in Eq.~(\ref{H}) 
the thermal Green's function (GF) ${\cal G}^{(0)}$ is given by 
\begin{equation}
{\cal G}^{(0)}_{\sigma}(i\nu_{n})=
\frac{1}{i\hbar \nu_{n}-\xi^{(0)}_{\sigma}}
\label{G0om}
\end{equation}
where $\nu_{n} = (2n+1)\pi/(\hbar\beta)$ is a fermion Matsubara
frequency, $\beta =  1/ k_B T$, and
$\xi^{(0)}_{\sigma} = - \sigma \Delta_z/2 - \mu$ ($\sigma=+1$ for 
$\uparrow$,  
$\sigma=-1$ for $\downarrow$) is the 
single-particle energy measured from the chemical potential.
In the strong magnetic field limit, translational
invariance implies\cite{densmat} not only that the Green's function is diagonal
in the momentum labels of the Landau gauge states, but also that the 
diagonal elements are independent of momentum.  This general property
leads to thermodynamic Green's functions, which depend only on Matsubara 
frequency and spin.  The Feynman rules for the evaluation of self--energy 
diagrams for the present problem are summarized in Appendix~\ref{feynman}.  

In the diagram sum we use to approximate the 
electronic self--energy in the following section. 
The propagators, which appear,  
are self-consistent Hartree-Fock (SHF) propagators rather 
than the bare propagators of Eq.~(\ref{G0om}).
The Hartree-Fock propagators are obtained by self-consistently solving the 
Dyson equation with the lowest order self-energy diagram 
illustrated in Fig. \ref{diag}.  This leads to an algebraic equation  
for the self-energy: 
\begin{equation}
\Sigma_{\sigma}^{HF} = \xi_{\sigma}^{HF} - \xi_{\sigma}^{(0)} = - \lambda \tilde{a}(0)
n_{F}(\xi_{\sigma}^{HF})\;. 
\label{SC}
\end{equation}
The SHF Green's function is
\begin{equation}
{\cal G}^{HF}_{\sigma}(i\nu_{n})=
\frac{1}{i\hbar \nu_{n}-\xi^{HF}_{\sigma}}\;.
\label{GF}
\end{equation}
Since the chemical potential at $\nu =1$ is determined by the equation
$n_F(\xi_{\uparrow})+ n_F(\xi_{\downarrow}) =1$ 
it follows that $\xi_{\uparrow}^{HF}(T)= - \xi_{\downarrow}^{HF}(T)<0$ and
that the chemical potential is fixed at $\mu = - \lambda \tilde{a}(0)/2$
{\em independent}  of temperature and Zeeman energy. For
$k_B T  <  \lambda \tilde{a}(0)/4$ and  
weak Zeeman coupling it can happen that Eq.~(\ref{SC}) has three solutions.
In this case we choose the lowest value of $\xi_{\uparrow}^{HF}$ since 
this is the solution, which minimizes the grand potential 
($\Omega = - 2 N k_B T  \ln(2 \cosh(\beta \xi_{\uparrow}/2)$). 
In Fig.~\ref{schf} we plot the Hartree-Fock eigenvalue $\xi_{\downarrow}^{HF}$ and the 
spin magnetization as a function of temperature.  The 
difference $\xi_{\downarrow}^{HF}-\xi_{\uparrow}^{HF} = 2
\xi_{\downarrow}^{HF}$ is the exchange enhanced spin
splitting \cite{AU74,UNHF90} of the lowest Landau level.  Its maximum value 
occurs at $T=0$ and is $ \tilde{a}(0)\lambda
+\Delta_z$, dominated by the interaction term for $\Delta_z \ll \lambda$. 
In the high temperature limit ($T \rightarrow \infty$) 
this gap reduces to the bare Zeeman splitting $\Delta_z$. \\

The sharp inflection point in the spin magnetization curve
for $\Delta_z=0.016 \lambda$ in Fig.~\ref{schf} is a remnant of
the spontaneous magnetization 
that occurs {\it incorrectly} in the SHF at low temperatures.  
For $\nu=1$ and $\Delta_z = 0 $ it follows from 
Eq.~(\ref{SC}) that $x \equiv \beta \xi_{\uparrow}^{HF}$ satisfies:
\begin{equation}
x = \beta \tilde{a}(0)\lambda \left (\frac{1}{2} -  \frac{1}{(e^x +1 )} \right ) \; .
\label{MFx}
\end{equation}
At high temperatures the only solution to this equation is $x=0$ so 
that $n_{F}(\xi_{\uparrow}^{HF})=n_{F}(\xi_{\downarrow}^{HF})=1/2$ and
there is no spin-polarization.  
Expanding the r.~h.~s.~of Eq.~(\ref{MFx}), we see that $x \ne 0$ solutions
are possible when the coefficient of the linear term exceeds one, 
{\em i.e.} for $T < T_c^{HF} = \tilde{a}(0) \lambda /(4 k_{B})$.
Expanding (\ref{MFx}) up to third order we find the expected 
mean-field behavior for $T$ near $T_c^{HF}$: 
\begin{equation}
\frac{M(T)}{M(T=0)} =
\sqrt{3} \frac{T}{T_c^{HF}} \left (1-\frac{T}{T_c^{HF}}\right )^{1/2}
\label{magx}
\end{equation}
Similarly, at $T_c^{HF}$ we obtain
\begin{equation}
\frac{M(T=T_c^{HF}, \Delta_z)}{M(T=0)}=
\frac{T}{T_c^{HF}} \left (\frac{3 \Delta_z}{\tilde{a}(0) \lambda}\right )^{1/3}
\end{equation}
with mean field exponent $\delta = 3$. 
The SHF spontaneous magnetization is plotted in Fig.~\ref{schf}.  

The finite $T$ ferromagnetic instability of the SHF also appears in the
the RPA expression for the spin susceptibility per area 
$\bar{\chi}^{-+}(\vec{q},i\omega_n) = 1/\hbar 
\int_{0}^{\hbar \beta} d\tau\; e^{i\omega_{n} \tau} \bar{\chi}^{-+}(\vec{q},\tau)$,
where  $\omega_{n}=2n\pi/(\hbar \beta)$ are bosonic Matsubara frequencies and 
\begin{equation}
\bar{\chi}^{-+}(\vec{q},\tau)=\frac{(g \mu_B)^2}{A} <T \bar{S}^{-}(\vec{q},\tau)
\bar{S}^{+}(-\vec{q},0)> \; .
\end{equation}
The overbar is intended to emphasize that the susceptibility is to 
be evaluated in the strong field limit where the Hamiltonian can 
be projected onto the LLL. 
The RPA expression can be obtained from a ladder diagram sum
with SHF Green's functions, similar to the sum for the self-energy
detailed in the next section.  The result is\cite{AM91}
\begin{equation}
\bar{\chi}^{-+(RPA)}(\vec{q},i\omega_n) = \frac{\bar{\chi}^{-+ (HF)}(\vec{q},i\omega_n)}
{1+I(\vec{q})\bar{\chi}^{-+(HF)}(\vec{q},i\omega_n)}
= -\frac{(g\mu_B)^2 e^{-q^2\ell_c^2/2} }{2 \pi \ell_c^2} 
\frac{(\nu_{\uparrow}^{HF}-\nu_{\downarrow}^{HF})}
{(i \hbar \omega_n-\tilde{\epsilon}_{SW}(\vec{q}))}
\label{susc}
\end{equation}
where the single-bubble HF spin-susceptibility is
\begin{equation}
\bar{\chi}^{-+(HF)}(\vec{q},i\omega_n) = 
\frac{(g\mu_B)^2 e^{-q^2\ell_c^2/2} }{2 \pi \ell_c^2} 
\frac{(\nu_{\uparrow}^{HF} - \nu_{\downarrow}^{HF})}
{(i \hbar  \omega_n - \xi_{\uparrow}^{HF} + \xi_{\downarrow}^{HF})}
\end{equation}
is the bubble with HF lines.  The effective interaction 
appearing in Eq.~(\ref{susc}) is 
defined by $I(\vec{q})= -2\pi \ell_c^2/(g \mu_B)^2 e^{q^2\ell_c^2/2}\tilde{a}(\vec{q}) 
\lambda$.
The quantity $\tilde{\epsilon}_{SW}(\vec{q})$ 
\\
\begin{equation}
\tilde{\epsilon}_{SW}(\vec{q})= \Delta_z + \lambda (\nu_{\uparrow}^{HF} -
\nu_{\downarrow}^{HF}) 
(\tilde{a}(0) - \tilde{a}(\vec{q}))
\label{SW1}
\end{equation}
reduces to the spin wave spectrum $\epsilon_{SW}(\vec{q})$ 
of Eq.~(\ref{DISP}) in
the $T \to 0$ limit.  In this approximation, 
the spin-wave bandwidth is reduced in proportion to the 
spin-polarization as the temperature increases.  Note that 
this approximation does not capture the finite lifetime
of spin-wave states, which will result from spin-wave 
spin-wave interactions at higher temperatures.  This will be
one of the important limitations of the theory we present in
the following section.  

For $\Delta_z =0$ the static limit of the RPA 
susceptibility, plotted in Fig.~\ref{schf}, diverges at the same  
temperature at which the spontaneous magnetization determined
by the SHF equations Eq.~(\ref{SC}) vanishes. 
These results are in disagreement with the Mermin-Wagner 
theorem, which forbids continuous broken symmetries at finite
temperatures in two dimensions.  The disagreement is expected for
mean-field theory.  It clear that
the SHF magnetizations calculated at the relatively small but finite
values of $\Delta_z$, appropriate for experimental systems,
will be too large. In essence, the SHF calculations of this section are 
equivalent to the Stoner band theory\cite{moriya,stoner}
of metallic ferromagnetism.  In both cases the ground
state is well described.  (In the present situation the SHF ground 
state and the exact ground state are identical.)
In both cases the magnetization 
is overestimated at finite temperatures primarily  
because of the failure to account for magnetization suppression
due to thermally excited collective spin-wave excitations. 
The approximation discussed in the following section remedies  
this gross deficiency.  The situation in the present  
two-dimensional systems with a small Zeeman coupling
is not unlike the situation in most three-dimensional 
metallic ferromagnets with no external field.  For example
it is generally accepted\cite{korenman}
that most but not all of the magnetization suppression
up to the critical temperature in the elemental metallic ferromagnets
(Fe, Ni, and Co) is due to spin-wave excitations. 
Of course, the separation of magnetization suppression into
collective spin-wave effects and the particle-hole effects
 cannot, in general, be made precise. 
In metallic ferromagnets, and in the present two-dimensional
systems, a complete theory valid at moderately high
temperatures requires the interplay between collective spin-fluctuations
and the underlying fermionic degrees of freedom to be accurately
described.  The next section reports on an adaptation of 
microscopic theories of metallic ferromagnets to the present case.

\section{Electron spin-wave scattering}

Our theory is based on an expansion in terms of SHF Green's functions.
The self-energy correction to these propagators can 
in principle be expressed in terms of the exact scattering 
vertex $\Gamma^{(4)}(1,2,3,4)$ and the exact GF. 
(Here, $1$ is short for $(q_{1}, i\nu_{1,n})$.) 
We use an approximation, which, as we shall demonstrate, captures 
much of the essential physics.  It is the analogue for quantum Hall
ferromagnets of the approximation discussed 
by Hertz and Edwards \cite{HE73} for the case of 
single-band Hubbard models with ground states, which are strong
ferromagnets.  The self-energy is approximated by a particle-hole 
ladder summation, which gives an exact description for interactions between
a single hole in the majority band and a single electron in 
the minority band.  However, as the density of spin-flip excitations
increases at higher temperatures, the approximation becomes 
less accurate. 

The Bethe-Salpeter integral equation for the 
scattering vertex is expressed diagrammatically in Fig.~\ref{leit}.
Explicit calculation of low order ladder-diagram 
particle-hole vertex parts using the
Feynman rules of Appendix~\ref{feynman} shows that 
in this approximation 
\begin{eqnarray}
\Gamma^{(4)}(1,2,3,4) = \Gamma^{(4)}(q+\Delta q/2 - q_{4},
\Delta q; i \omega_{n})
\end{eqnarray}
depends only on two momenta and one frequency.  Here,   
$q=(q_1 + q_3)/2$ and $\Delta q = q_1 - q_3$ are the center of mass 
and the relative momenta, respectively, and $i \omega_{n} = 
i( \nu_{3,n} - \nu_{1,n})$ is the bosonic Matsubara frequency of
the relative motion of the ingoing and outgoing particles. 

The Bethe-Salpeter integral equation for the 
scattering vertex is 
\begin{eqnarray}
\Gamma_{\sigma,\sigma'}^{(4)}(q+ \Delta q/2 - q_4,\Delta q; & &  i \omega_n)  = 
\lambda \tilde{W}(q+ \Delta q/2 - q_4,\Delta q)  \nonumber \\ 
  + & &  \lambda \bar{ \chi}_{\sigma,\sigma'}(i \omega_n)
\ell_c \int_{-\infty}^{\infty} dq' \tilde{W}(q-q',\Delta q)
\Gamma_{\sigma,\sigma'}^{(4)}(q'+ \Delta q/2 - q_4,\Delta q; i \omega_n) . 
\label{BS}
\end{eqnarray}
Here, we have introduced the pair propagator, 
$\bar{\chi}_{\sigma,\sigma'}(i \omega_n)$, defining 
\begin{equation}
\bar{\chi}_{\sigma,\sigma'}(i \omega_n) = - \frac{1}{\beta}
\sum_{i \nu_{n}} {\cal G}_{\sigma}(i \nu_{n}) 
{\cal G}_{\sigma'}(i (\nu_{n} + \omega_{n})).
\label{pp}
\end{equation}

It is important to note that the $q$-independence of the GF in the 
case of quantum Hall ferromagnets immensely simplifies the solution 
of the Bethe-Salpeter equation.  
In fact, Eq.~(\ref{BS}) can be reduced to an algebraic equation  
because the second term on the RHS is a convolution integral in the 
center of mass coordinate\cite{KH84}.   
We define a partial Fourier transformation of the scattering function: 
\begin{eqnarray}
\tilde{\Gamma}^{(4)}_{\sigma,\sigma'}(p,\Delta q;i \omega_n) 
&  =&  \ell_c \int_{-\infty}^{\infty} dq e^{ipq\ell_c^2}\Gamma_{\sigma,\sigma'}^{(4)}(q,\Delta q;i \omega_n) \nonumber \\
& = & e^{ip(\Delta q/2 - q_4)\ell_c^2} \ell_c  
\int_{-\infty}^{\infty} dq e^{ipq \ell_c^2}\Gamma^{(4)}_{\sigma,\sigma'}(q+ \Delta q/2 - q_4,\Delta q; i \omega_n),  
\label{FT}
\end{eqnarray}
The corresponding transformation of the Landau-gauge vertex
$\tilde{W}(q,\Delta q)$ is the  
particle-hole interaction, which appeared previously in Eq.~(\ref{SW1}): 
\begin{eqnarray}
\tilde{a}(\vec{k}=(\Delta q, p)) = 
\ell_c \int_{-\infty}^{\infty} dq e^{ipq\ell_c^2} \tilde{W} (q,\Delta q) = 
\int \frac{d^{2}\vec{q'}}{(2\pi)^2} 
\tilde{V}(\vec{q'})e^{-q'^2\ell_c^2/2} e^{i(q'_{x}\Delta q+q'_{y}p)\ell_c^2} \;, 
\label{FTvertex}
\end{eqnarray}
where we have defined the two-dimensional wavevector $\vec{k}= (\Delta q, p)$. 
With these definitions we find that 
\begin{eqnarray}
\tilde{\Gamma}^{(4)}_{\sigma,\sigma'}(\vec{k}=(p,\Delta q);i \omega_n)
= \frac{\tilde{a}(\vec{k})\lambda}{(1 - \bar{\chi}_{\sigma,\sigma'}(i \omega_n) \tilde{a}(\vec{k})\lambda)}. 
\label{solution}
\end{eqnarray}

Our self-energy approximation consists of combining this 
scattering vertex with the propagators.
In order to avoid double counting the first order terms
already present in the Hartree-Fock propagators, we must
subtract the first order term from the r.~h.~s.~of Eq.~(\ref{solution}).
Inverting the transform of Eq.~(\ref{FT}) then gives the 
sum of second and higher order terms in the Bethe-Salpeter
equation.  The approximate self-energy for spin $\sigma$ 
is obtained by contracting the incoming and outgoing lines
of the opposite spin into a Hartree-Fock propagator.  
The result is that 
\begin{eqnarray}
\tilde{\Sigma}_{\sigma}(i \nu_{n}) = \frac{1}{\beta} \sum_{i \omega_{n}}
{\cal G}_{\sigma'}
(i (\nu_{n}+\omega_{n})) \tilde{\Gamma}^{(4)}_{\sigma,\sigma'}(i \omega_n) \; ,
\label{SE}
\end{eqnarray}
where 
\begin{eqnarray}
\tilde{\Gamma}_{\sigma,\sigma'}^{(4)}(i \omega_n) \equiv  2 \pi \ell_c^2 \int \frac{d^2 \vec{k}}{(2 \pi)^2} 
\tilde{\Gamma}^{(4)}_{\sigma,\sigma'}(\vec{k};i \omega_n) 
= 2 \pi \ell_c^2 \int \frac{d^2 \vec{k}}{(2 \pi)^2} \biggl\lbrace
\frac{\tilde{a}(\vec{k}) \lambda}
{(1 - \bar{\chi}_{\sigma,\sigma'}(i \omega_n) \tilde{a}(\vec{k})\lambda)}   - \tilde{a}(\vec{k})\lambda \biggr\rbrace . 
\label{Gamma}
\end{eqnarray}

The Dyson equation relating the SHF and full  
Green's function is 
\begin{equation}
({\cal G}_{\sigma}(i\nu_{n}))^{-1} - ({\cal G}^{HF}_{\sigma}(i\nu_{n}))^{-1}
= - \tilde{\Sigma}_{\sigma}(i \nu_{n}).
\label{Dyson}
\end{equation}
We address below the question of whether the SHF Green's function
or the corrected Green's function should be used in the 
expression for the pair propagator.
If we use SHF Green's functions, the particle-hole ladder diagram
of order $n$ is proportional to $(\nu_{\sigma}^{HF}-\nu_{\sigma'}^{HF})^{(n-1)}$.
At low temperature this factor is close to one and larger than
the combinations of filling factors, which result from other diagrams
of the same order. {\em E. g.}, the corresponding
particle-particle ladder is proportional 
to $(1 -\nu_{\sigma}^{HF}-\nu_{\sigma'}^{HF})^{(n-1)}$, which is zero 
for any $T$ when $\sigma \ne \sigma'$.  
This observation may be used at low temperatures 
to systematically justify the class of diagrams we have included.
At higher temperatures we expect that the sums of ring diagrams, which
describe screening physics\cite{SMG92} will be among the 
important omissions.  

It is possible to require that the Green's functions used
in the pair-propagator and self-energy
expressions be obtained from the Dyson equation.
This leads to a set of coupled equations, which can 
be solved numerically.  For this purpose it is 
convenient\cite{Hau96} to express the equations in the following 
mixed imaginary-time imaginary-frequency representation:
\begin{eqnarray}
{\cal G}^{-1}_{\sigma}(i\nu_{n}) - ({\cal G}^{HF}_{\sigma}(i\nu_{n}))^{-1}
& = & - \tilde{\Sigma}_{\sigma}(i \nu_{n}) \nonumber \\
\tilde{\Sigma}_{\sigma}(\tau) & = & \tilde{\Gamma}_{\sigma, \sigma'}^{(4)}(- \tau) {\cal G}_{\sigma'}( \tau)
\nonumber \\
\tilde{\Gamma}_{\sigma,\sigma'}^{(4)}(i \omega_n) 
& = &
2 \pi \ell_c^2\int \frac{d^2 \vec{k}}{(2 \pi)^2} \biggl\lbrace \frac{\tilde{a}(\vec{k})\lambda}
{(1 - \bar{\chi}_{\sigma,\sigma'}(i \omega_n) \tilde{a}(\vec{k})\lambda )}   - \tilde{a}(\vec{k}) \lambda \biggr\rbrace 
\nonumber \\ 
\bar{\chi}_{\sigma,\sigma'}(\tau) & = & - {\cal G}_{\sigma}(-\tau) {\cal G}_{\sigma'}(\tau)
\label{system}
\end{eqnarray}
As we discuss later, 
we have found that these self-consistent equations tend not to have
stable solutions at low temperature.

Our work is based mainly on 
the approximation in which the pair propagator and 
the self-energy are evaluated with SHF Green's functions.
The pair-propagator frequency sum may then be evaluated analytically: 
\begin{eqnarray}
\bar{\chi}_{\sigma,\sigma'}^{HF}(i \omega_n) = -
\frac{(n_{F}(\xi_{\sigma}^{HF}) 
- n_{F}(\xi_{\sigma'}^{HF})) }{( i \hbar \omega_n +  \xi_{\sigma}^{HF} - \xi_{\sigma'}^{HF}) }\; ,
\label{pp1}
\end{eqnarray}
leading to the following explicit expressions for the corrections $\tilde{\Sigma}_{\sigma}(i\omega_{n})$ of
the majority spin and minority spin SHF self-energies: 
\begin{eqnarray}
\tilde{\Sigma}_{\uparrow}(i\nu_{n}) = \lambda^{2}(\nu_{\uparrow}^{HF} - \nu_{\downarrow}^{HF})
\int_{0}^{\infty}
d(\frac{k^2\ell_c^2}{2}) \tilde{a}^{2}(k)
\frac{\lbrace n_{B}(\tilde{\epsilon}_{SW}(k)) + n_{F}(\xi_{\downarrow}
^{HF})\rbrace }{(i \hbar \nu_{n} + \tilde{\epsilon}_{SW}(k) -
\xi_{\downarrow}^{HF})}
\label{SE1}
\end{eqnarray}
and
\begin{eqnarray}
\tilde{\Sigma}_{\downarrow}(i\nu_{n}) =  \lambda^{2}(\nu_{\uparrow}^{HF} - \nu_{\downarrow}^{HF})
\int_{0}^{\infty}
d(\frac{k^2\ell_c^2}{2}) \tilde{a}^{2}(k)
\frac{\lbrace n_{B}(\tilde{\epsilon}_{SW}(k)) + 1 - n_{F}(\xi_{\uparrow}
^{HF}) \rbrace}{(i\hbar \nu_{n} - \tilde{\epsilon}_{SW}(k) - \xi_{\uparrow}^{HF})}
\; .
\label{SE2}
\end{eqnarray}
Here, 
$n_{B}(\tilde{\epsilon}_{SW}(k))$ is the Bose-Einstein distribution function 
for the spin-waves whose dispersion is specified in Eq.~(\ref{SW1}). 
For $T=0$ the occupation factors in the numerators  
of both self-energy expressions vanish, and the  
SHF-result, which is exact in this limit, is recovered. 

These electronic self-energy expressions resemble those due to 
virtual phonon exchange in an electron-phonon system.\cite{Mah90}
The majority spin self-energy includes contributions
from processes where a majority spin electron scatters out to a 
minority spin state upon absorption of a spin-wave and processes 
where a minority spin electron scatters into a majority spin state 
upon emission of a spin-wave.  Because the spin-wave carries spin
$S_z = -1$ there are no processes where a majority spin electron 
scatters to a minority spin state and emits a spin-wave or a minority 
spin electron scatters to a majority spin state and absorbs a spin-wave.
This distinction explains both the difference between the 
majority spin self-energy in Eq.~(\ref{SE1}) and the 
minority spin self-energy in Eq.~(\ref{SE2}) and the difference between 
these self-energies and the phonon exchange self-energy where
both classes of contribution appear at once\cite{Mah90}.  
The electron self-energy expressions are identical to those,  
which would be obtained for a model where the electrons and
spin-waves were regarded as independent fermion and boson particles 
with an interaction in which fermions are scattered by 
emitting or absorbing spin-waves.  The effective electron spin-wave
interaction, which can be read-off from the self-energy expressions,
is proportional to $\lambda \tilde a(\vec k)$.  Note that, unlike 
the case of deformation potential electron-phonon coupling, the 
matrix element approaches a constant as $k \to 0$.  
If the long-range  Coulomb interaction is not screened, 
the electron spin-wave interaction 
falls off only as $|k|^{-1}$ for large $k$.
(For ideal 2D Coulomb interactions 
$\tilde a(\vec{k}) = \sqrt{\pi/2}e^{-k^2 \ell_c^2/4}I_{0}(k^2 \ell_c^2/4)$.) 
The electron spin-wave interaction at large $k$ can be understood
in terms of the excitonic picture discussed in Section II and is 
proportional to the electron-electron interaction at {\it real space} 
separation $ k \ell_c^2$.  If the 
Coulomb interaction is screened $\tilde a(\vec k)$ will
begin to fall off more quickly once $k \ell_c^2$ exceeds $\sim k_{sc}^{-1}$.
The slow fall off of these large 
momentum transfer scattering events requires us to 
take account of screening when we evaluate the self-energy
expressions.

Note that these self-energy expressions satisfy the equation  
$\tilde{\Sigma}_{\uparrow}(i\nu_n) =-\tilde{\Sigma}_{\downarrow}(-i\nu_n)$. 
This is an exact identity for the case $\nu = 1$, which
follows from the particle-hole symmetry\cite{kunyang} of the 
underlying Hamiltonian.
We remark that the self-energies given by Eq.~(\ref{SE1}) and 
Eq.~(\ref{SE2}) have branch cuts along a finite 
portion of the real line.  For majority 
spins the branch cut occurs along the interval 
$I_{\uparrow} = (\xi^{-}_{\uparrow}=\xi^{HF}_{\uparrow},  
\xi^{+}_{\uparrow}=\xi^{HF}_{\downarrow} - \Delta_z)$.
For the minority-spin the branch cut interval is 
$I_{\downarrow}= (\xi^{-}_{\downarrow}=\xi^{HF}_{\uparrow}+ \Delta_z, 
\xi^{+}_{\downarrow}=\xi^{HF}_{\downarrow})$).
Outside of these intervals the self-energy is real on the real line.
Because of the branch cuts, some care
is required in the numerical evaluation of the self-energy 
expression.

In this paper we concentrate on physical properties which can
be expressed in terms of the one-particle real-time Green's
function.  Analytically continuing
the thermal Green's function self-energy expressions,  
Eqs.~(\ref{SE1}) and (\ref{SE2}), to the 
real frequency axis ($i \hbar \nu_n \rightarrow E + i \eta$)
gives $ G^{ret}_{\sigma}(E)=1/(E + i \eta - \xi^{HF}_{\sigma}-
\tilde{\Sigma}_{\sigma}(E))$.
The retarded Green's function is completely specified by its spectral
function 
\begin{equation} 
A_{\sigma}(E) = - 2 Im G^{ret}_{\sigma}(E).
\label{SF1}
\end{equation}
It is $A_{\sigma}(E)$, which we evaluate numerically, and 
we start by mentioning some of its general properties.
If we consider the system of equations defining the 
SHF-GF (\ref{system}) and start the iteration from
the bottom equation 
with a GF satisfying ${\cal G}_{\uparrow}(i \nu_n) = 
- {\cal G}_{\downarrow}(-i  \nu_n)$, 
we end up with a new GF satisfying 
the same relation, {\it i.e.} the approximate system of equations 
conserves this property.  The SHF-GF has this property since 
$\xi^{HF}_{\downarrow} = - \xi^{HF}_{\uparrow}$.
Since $G_{\uparrow}^{ret}(E)= - G_{\downarrow}^{av}(-E)$, 
the spectral functions for up and down spins satisfy the following 
relationship:
\begin{equation}
A_{\uparrow}(E) = A_{\downarrow}(-E). 
\label{SF2}
\end{equation}
Therefore the knowledge of the spectral function for only one spin direction 
for the case of $\nu = 1$ is sufficient to determine the result for the other 
spin direction.  

When the self-energy is evaluated from the Hartree-Fock GF 
the qualitative behavior of $A_{\uparrow}(E)$ 
can be understood from the analytical structure of the denominator 
$(E  - \xi^{HF}_{\uparrow} - \tilde{\Sigma}_{\uparrow}(E))$
of the GF. 
The spectral function for $\sigma = \uparrow$ is non-zero along the branch cut 
where the retarded self-energy has a non-zero imaginary part,
{\em i.e.} for $\xi^{HF}_{\uparrow} \le E \le
\xi^{HF}_{\downarrow}-\Delta_z$.
Since the real part of the self-energy is monotonically decreasing
outside of this interval, it vanishes for $E \to \pm \infty$, 
and is divergent for $E \to
\xi^{HF}_{\downarrow}-\Delta_z$ from above (see below),
it follows that the Green's function also has simple poles,
and the spectral function has 
$\delta$-function contributions on both sides of the branch cut.
The positions of these quasiparticle poles in the Green's function 
are determined by 
\begin{equation}
E -\xi_{\sigma}^{HF} =
\tilde{\Sigma}_{\sigma}^{ret}(E) \;.
\label{zero}
\end{equation} 
Fig.~\ref{real} illustrates a graphical solution of 
this equation for $T = 0.1 \lambda / k_B$.
As we discuss below, the spectral weight is dominated by the
two $\delta$-function contributions at $E_{\sigma}^{-}$ and 
$E_{\sigma}^{+}$ 
except at elevated temperatures. 

To get a qualitative feel for the physics of the self-energy 
at low temperatures, 
it is useful to divide it into separate contributions from 
interactions with collective long wavelength spin-waves and 
from interactions with short wavelength
spin-$\downarrow$ electron,  spin-$\uparrow$ hole  
pairs.  We arbitrarily treat spin-waves with $ k \ell_c < 1$ as 
collective and those with $k \ell_c > 1 $ as single-particle.
For collective spin-waves we can approximate the 
electron-spin-wave interaction by a constant $\lambda \tilde a (0)$ 
and the low-energy spin-wave dispersion by $\Delta_z + 4 \pi \rho_s \ell_c^2 k^2$.
Similarly, for the particle-hole states we can, when 
screening is neglected, approximate the interaction by $\lambda / k \ell_c $ 
and the spin-wave energy by $ 2 \xi^{HF}_{\downarrow} - \lambda / k
\ell_c$.  We first concentrate on the region of energy $E$ near 
$\xi^{HF}_{\uparrow}$ where the main spectral weight resides at 
low temperatures.  The collective spin-wave contribution to the 
self-energy near the low energy quasiparticle is
\begin{equation}
\tilde \Sigma^{C-}_{\uparrow}(E) 
\approx - (\lambda \tilde a (0) \ell_c)^2 \int_0^{\ell_c^{-1}} dk k
\frac{n_B(\epsilon_{SW} (k))}{\lambda \tilde a (0) +
\xi_{\uparrow}^{HF} -E}.
\label{eq:cll}
\end{equation} 
We have assumed here that the spin-wave Bose factor is 
larger than spin-down electron Fermi factor.  Note that this 
requires, at a minimum, that $\Delta_z$ be less than 
{\it half} the single-particle energy
gap $\xi_{\downarrow}^{HF} - \xi_{\uparrow}^{HF}$. 
The extra factor of two in this condition occurs because of the itinerant 
nature of the single-particle excitations.  This self-energy
contribution will be negative and only weakly energy-dependent
for $E < \xi^{HF}_{\uparrow}$.  The particle-hole 
contribution to the self-energy in the same energy range is
\begin{equation}
\tilde \Sigma^{PH-}_{\uparrow}(E)
\approx - \lambda^2 \int_{\ell_c^{-1}}^{\infty} d k k^{-1} 
\frac{n_F(\xi_{\downarrow}^{HF})}{\xi_{\uparrow}^{HF} + \lambda / k \ell_c
- E}.
\label{eq:phl}
\end{equation}
This contribution to the self-energy is sharply energy dependent,
reaching a maximum for $E = \xi_{\uparrow}^{HF}$, and is 
formally divergent for every energy.  The divergence comes from
the large $k$ contribution to the integral, {\it i.e.} from  
the interaction of electrons with widely separated spin-flip
electron-hole pairs.  It can be cured by introducing screening into
our theory as discussed earlier.  Screening effectively cuts off 
the divergent integral at $k \sim 1/ k_{sc} \ell_c^2$.   
At low $T$, a Thomas-Fermi approximation estimate
would give $k_{sc} \ell_c \sim \lambda 
\exp ( - \xi^{HF}_{\downarrow}/ k_B T) $. The screening
wavevector is exponentially small because of the gap for 
charged excitations and the ultraviolet cutoff is 
consequently exponentially large.  However, except at 
$E = \xi^{HF}_{\uparrow}$, the particle-hole  
self-energy contribution depends only logarithmically on the 
screening vector and the declining Fermi factor will result
in a small contribution at low temperatures.  At $E =
\xi^{HF}_{\uparrow}$, the particle-hole contribution depends 
linearly on $k_{sc}^{-1}$ leading to a very narrow peak 
near this energy at low temperatures.    

Because of the collective contribution to the self-energy, 
the low-energy quasiparticle pole will be shifted to 
energies below $\xi^{HF}_{\uparrow}$ and away from
this peak.  In the end the particle-hole contribution to the
self-energy is much larger than it would be for a system
with short range interactions.  Nevertheless, provided that 
the collective gap $\Delta_z$ is much less than 
half the particle-hole gap, there will be a region at low-temperatures
where its contribution becomes unimportant.  Neglecting this
contribution we find the low-energy pole has residue 
\begin{equation} 
 z = \big[ 1 + \ell_c^2  \int_0^{\ell_c^{-1}} dk k n_B (\epsilon_{SW} (k))
\big]^{-1}
\label{zlow}
\end{equation} 
and occurs at energy 
\begin{equation} 
E_{\uparrow}^- = \xi^{HF}_{\uparrow} - \lambda \tilde{a}(0) [ 1 -z ] 
\label{elow}
\end{equation} 
The electron-spin-wave interaction strength drops
out of the expression for $z$ because of its relationship
to the spin-flip excitation energies. 
As we discuss below, the loss of spectral weight due to occupied 
spin-waves gives a magnetization suppression identical to that which
would be obtained from a non-interacting spin-wave model with 
appropriate ultraviolet cutoffs.

A similar calculation can be carried out for the high 
energy pole in the Green's function.  For $E =
\xi^{HF}_{\downarrow} - \Delta_z + \delta $ and $\delta$ small 
we can again identify approximate 
collective and particle-hole contributions to the self-energy:
\begin{equation}
\tilde \Sigma_{\uparrow}^{C+}(E) \approx (\lambda \tilde a (0) \ell_c )^2 \int_0^{\ell_c^{-1}} dk k
\frac{ n_B (\epsilon_{SW}(k))}{ \delta + 4 \pi \rho_s k^2 \ell_c^2 }
\label{sigcolp}
\end{equation}
and 
\begin{equation}
\tilde \Sigma_{\uparrow}^{PH+}(E) \approx \lambda  \int_{\ell_c^{-1}}^{\infty} dk k^{-1}
\frac{ n_F(\xi^{HF}_{\downarrow})}{ \tilde a(0) }. 
\label{sigphp}
\end{equation}
In this case, screening is not necessary to make the particle-hole
contribution small at low temperatures.  The second term in
the denominator of Eq.~(\ref{sigcolp}) is negligible if 
$\delta \ge k_B T $ because of the exponential cut-off in the 
spin-wave Bose occupation factors; we see below that condition
is satisfied at the upper pole in the Green's function.
Keeping only the collective contribution gives a high energy
pole with residue $z^{+} = z^{-1}-1$ and energy
\begin{equation}
E_{\uparrow}^{+} = \xi^{HF}_{\downarrow}-\Delta_z + \lambda \tilde a (0) [z^{-1}-1].
\end{equation} 
For $4 \pi \rho_s >>  k_B T >> \Delta_z $, it 
follows from Eq.~(\ref{zlow}) that 
\begin{equation}
z^{-1}-1 =  \frac{k_B T}{ 8 \pi \rho_s} \ln (k_B T / \Delta_z).
\end{equation} 
which, since $16 \pi \rho_s = \lambda \tilde a(0)$, guarantees
that $\delta$ is in the assumed energy range.  The approximations
made above fail when $z^{-1}-1$ is large, in which case the 
sums of the approximate residues exceeds one.  When $1-z$ is small, 
it follows from these calculations that the spectral weight of 
the Green's function is exhausted by the two poles in the 
Green's function.  As spin-waves are excited, weight is shifted
from the low energy pole, which is below the chemical potential,
to the high-energy pole, which is above the chemical potential.
In the limit of zero-temperature, the spectral weight 
lies entirely in the low-energy pole and Hartree-Fock theory
results are recovered. 

As we see from Fig.~\ref{real} and from the above discussion,
the real part of the self-energy diverges to $ - \infty$ at the 
lower limit of the branch cut when screening is neglected 
and to $ + \infty $ at its upper limit, even if screening is 
included.
The divergence at $E=\xi_{\uparrow}^{HF}$
in Fig.~\ref{real} produces only
a small feature in our numerical calculations because 
these results were obtained with $k_{sc} \ne 0$.
However, the divergence at the upper boundary of the branch cut
is clearly visible. 
The imaginary part of the GF is nonzero throughout 
the interval from $\xi_{\uparrow}^{HF}$ to 
$\xi_{\downarrow}^{HF}-\Delta_z$ because spin-flip
excitations exist at all energies between $\Delta_z$ and 
$\Delta_z + \lambda \tilde a (0)$. 
The imaginary part of the retarded Green's function is negative definite,
as illustrated in Fig.~\ref{imag}. 

The explicit expression for the spectral density is 
\begin{eqnarray}
A_{\sigma}(E) & = & \frac{2\pi \delta(E-E_{\sigma}^{-})}
{|1 -  \frac{\partial \tilde{\Sigma} 
{\sigma}^{ret}(E)}{ \partial E}_{|E = 
E_{\sigma}^{-}}|} +
\frac{2\pi \delta(E-E_{\sigma}^{+})}
{|1 -  \frac{\partial \tilde{\Sigma}
{\sigma}^{ret}(E)}{ \partial E}_{|E = E_{\sigma}^{+}}|}
\nonumber \\
&+& \theta(E-\xi^{-}_{\sigma})\theta(\xi^{+}_
{\sigma}-E)
\frac{(-2 Im \tilde{\Sigma}_{\sigma} ^{ret} (E))}{((E-\xi_{\sigma}^{HF}
-Re \tilde{\Sigma}_{\sigma}^{ret}(E))^2 + 
(Im \tilde{\Sigma}_{\sigma}^{ret} (E))^2)} \; . 
\label{Spectral}
\end{eqnarray}
The condition that the integral of 
the spectral function over frequency equals one 
can be used to check the accuracy of our calculations.

We have previously\cite{KM96} published figures illustrating the 
spectral densities obtained from numerical evaluations of this 
self-energy expression at several temperatures.  In agreement with the 
preceding analysis the spectrum consists of low energy and 
a high energy delta functions separated by a band associated
with the branch cut of the self-energy.  The incoherent band
contribution to the spectral weight tends to be peaked 
toward its high energy extremum where the imaginary part of
the self-energy has contributions from long-wavelength spin-waves. 
The spectral weight
shifts with increasing temperature from the low-energy Hartree-Fock pole
to the high-energy pole and partially to the intermediate 
energy band.  The spectral weight shift can be understood in 
terms of a reduction in the probability that majority spins will
be lined up with the exchange field from the ordered moment,
which fluctuates when spin-waves are excited at finite temperatures.   
In Fig.~\ref{weights} we plot the fraction of the 
spectral weight coming from these three contributions as a function
of temperature.  Notice that the incoherent band contribution
grows rather slowly with temperature.  In Fig.~\ref{positions}
we have also plotted the positions of the two poles 
band as a function of temperature.  Initially the two poles are 
separated by the zero-temperature exchange splitting gap.
In the approximation we employ, the splitting increases at
finite temperatures because of 
level repulsion with the continuum states.  
The results shown in Fig.~\ref{positions} are for 
screened interactions between the electrons; accounting for
the finite width of the quantum well or heterojunction reduces
the splittings to approximately $2/3$ of these values for 
typical systems and screening associated with Landau 
level mixing will cause a further reduction.

In the next section we discuss several experiments, which can test
the predictions which result from this spin-wave exchange
approximation for the self-energy.  Potentially the most 
telling of these will be 2D-2D tunneling experiments which
measure the spectral functions fairly directly.  We do not
expect perfect agreement between the present theory and experiment,
although qualitative agreement seems certain.  We expect that 
comparison with tunneling experiments will assist future
theoretical progress.  Tunneling spectrum measurements in 
typical band ferromagnets are much less informative because 
the band width (which is zero in the present problem) is
comparable to the quasiparticle band spin-splitting.\cite{prljmr}

\section{Observables} 

\subsection{The spin magnetization} 

The spin-magnetization is proportional to the difference of the 
occupation probabilities for spin-up and spin-down electrons,
$M(T) = M_0 (\nu_{\uparrow}- \nu_{\downarrow})$ where $M_0 
= N |g \mu_B| /2$ is the ground state spin-magnetization.  
Expressing the occupation probability, $\nu_{\sigma}$, 
in terms of the spectral functions gives: 
\begin{eqnarray}
\frac{M(T)}{M_0} &=&
\int_{- \infty}^{\infty} \frac{d E} {2 \pi} 
n_{F}(E)(A_{\uparrow}(E)- A_{\downarrow}(E))  \nonumber \\ 
 &=& - \int_{- \infty}^{\infty} \frac{ d E}{2 \pi}
  \tanh (\beta E /2)  A_{\uparrow}(E)
\label{M3}
\end{eqnarray}
For non-interacting electrons
$A_{\uparrow}(E) = 2 \pi \delta (E + \Delta_z/2)$ and 
$M/M_{0} = \tanh(\beta \Delta_{z}/4)$.  Since interactions tend to 
favor parallel spin alignment, we expect this result to be 
a lower bound for $M/M_0$.   As discussed in more
detail later, it should become accurate in both high and 
low temperature limits.  Note that it reflects the itinerant
nature of the electrons which carry the spin-magnetization.  
For localized spin-half particles 
$M/M_0 = \tanh(\beta \Delta_{z}/2)$; the magnetization is smaller 
at high temperatures in the itinerant case because of the 
number of many-particle states increases more 
rapidly with the number of reversed spins.  
In Hartree-Fock theory, $A_{\uparrow}(E) 
= 2 \pi \delta (E - \xi_{\uparrow}^{HF})$ and 
$M/M_0= \tanh(\beta \xi_{\downarrow}^{HF}(\beta)/2)$.
This prediction for the spin-magnetization is illustrated in 
Fig.~\ref{magn4}.  Because of the exchange enhanced spin-splitting 
the magnetization is much larger at fixed temperature
than in the noninteracting case.  
This result grossly overestimates the magnetization because,
as\cite{korenman} in the band theory of metallic magnetism, 
magnetization suppression due to thermally excited spin-waves 
is not accounted for.  These two simple results for $M/M_0$ 
should therefore bound the exact result.

Now let us turn to the results obtained using our self-energy
approximation which accounts for electron-spin wave scattering; 
numerical results obtained using two different values of 
$k_{sc}$ are shown in Fig.~\ref{magn4}.  As explained in the 
previous section, even though the screening wavevector must
be finite to ensure convergence of the wavevector integrals in our 
self-energy expressions, we do not expect great sensitivity
to its value until it becomes comparable to $l_c^{-1}$.
We have estimated the appropriate value for $k_{sc}$ 
at two different temperatures as described below. 
For $T\sim 0.09 \lambda/k_B$ and $T\sim 0.18 \lambda /k_B$, we 
find that $k_{sc}=0.01 l_{c}^{-1}$ 
and $k_{sc}=0.1l_{c}^{-1}$, respectively. 
The screening becomes weaker at low temperatures as the system
approaches incompressibility.  The magnetization curves of 
Fig.~\ref{magn4} were calculated with these two fixed screening 
wavevectors;
if the screening wavevector were allowed to be temperature dependent 
the magnetization value would be above the long-dashed curve in 
Fig.~\ref{magn4}
for temperatures below 0.09 $\lambda/k_B$
in Fig.~\ref{magn4}, should interpolate between the long-dashed 
short-dashed curves temperatures between $0.09 \lambda/k_B$ and $0.18
\lambda/k_B$, and below the short--dashed curve 
for temperatures beyond $0.18 \lambda/k_B$. 
The comparably weak dependence of the magnetization on the  
screening wavevector is expected, limiting this source of 
uncertainty in our predictions.
Fig.~\ref{magn4} shows that, for $\Delta_z = 0.016 \lambda$ 
the magnetization decreases almost linearly with $T$ over a wide 
range of temperatures  between $ \sim 0.01 \lambda/k_B$ and
$ \sim 0.2 \lambda/k_B$.  Over this temperature range, the 
portion of the spectral weight at $E < 0$ is dominated 
by the low-energy pole, and the Fermi factor evaluated at this 
pole is still close to 1.  Under these conditions the temperature
dependence comes nearly entirely from that of the renormalization factor at the 
low energy pole so that $M/M_0 \sim  (2 z -1)  \sim 1 - (k_B T) \ln
(k_B T/\Delta_z)/4 \pi \rho_s$.   
This temperature dependence is 
identical to what would be obtained from a non-interacting 
spin-wave model in an external magnetic field $B=\Delta_z/|g \mu_B|$.  
Ignoring the logarithmic factor, this 
effect is sufficient to reduce $M$ to small values for 
$k_B T \sim 4 \pi \rho_s \sim 0.3 \lambda $, and that is roughly what 
we observe in Fig.~\ref{magn4}.  

The accuracy of our calculation of spin-magnetization values 
is most reliably judged by comparing with results obtained 
by exact diagonalization of the many-particle Hamiltonian
for a small number of electrons on a sphere.\cite{CP96}
In Fig.~\ref{finsize} we present results for the ideal
Coulomb interaction obtained 
for $N=7$, and $N=9$ electrons on a sphere and compare them
with the results obtained from our self-energy approximation.
We can conclude from this comparison that our self-energy 
approximation overestimates the magnetization by approximately a
factor of two at intermediate temperatures.  We discuss
the physics behind this behavior at greater length below.
The magnetization values are least accurate at temperatures
between $\sim \Delta_z$ and 
$\sim 0.1 \lambda $ where finite-size effects, discussed 
below, have some importance and cause the magnetization per particle
to be underestimated by finite-size calculations.
Nevertheless, it seems clear that our 
simple self-energy expression results is an overestimate of the 
magnetization for $0.1 \lambda < k_B T < 0.3 \lambda $.
It is interesting to compare these results with essentially  
exact results for a $S=1/2$ Heisenberg
model on a square lattice\cite{timm} with a nearest
neighbor exchange interactions
whose strength has been adjusted to reproduce the spin-stiffness
of the quantum Hall ferromagnet.
(For the Coulomb model the spin stiffness is known exactly and has the value 
$\rho_s = \lambda \sqrt{\pi/2}/ 16 \pi$.)  
For the magnetization, this model is accurate when long
length scale collective magnetization degrees of freedom 
have dominant importance.  
For $k_B T$ smaller than $\sim 0.1 \lambda$, the Heisenberg
model magnetizations are slightly larger than those of the $N=9$ 
exact diagonalization calculations, presumably because they have
smaller finite-size corrections and are therefore more 
accurate, see Fig.~\ref{heisenberg}.   However, the microscopic and Heisenberg model 
$M(T)$ curves are very similar for 
$0.1 \lambda < k_B T < 0.2 \lambda$. 
The localized spin model $M(T)$ must fall below 
the exact diagonalization results for the electron model
at sufficiently high temperatures,
where the Heisenberg model magnetization is twice
as large as the magnetization of the itinerant electron system,
but this apparently does not occur until 
larger temperatures are reached. 
This figure also shows the results obtained\cite{timm}
by Timm {et al.} by evaluating leading $1/N$ corrections to
the magnetizations of continuum $SU(N)$ and $O(N)$ models, extending 
earlier work by Read and Sachdev.\cite{RS95}  
At moderate temperatures, the results obtained using
these approximation schemes are numerically better than
those obtained with our self energy approximation;
they are also somewhat unsatisfactory, however, since 
the $SU(N)$ scheme leads to negative magnetizations at 
high temperatures and the $O(N)$ scheme fails to 
capture the low-temperature non-interacting spin-wave limit.
Moreover, the continuum field theory approach cannot 
address important microscopic electronic properties like
the tunneling density of states. 
We discuss the experimental values of $M(T)$, also shown in this
figure, at greater length below.

Our exact diagonalization results for the magnetization 
have the largest finite-size errors when $\Delta_z$ is 
small.  In order to examine the limit of small $\Delta_z$ we 
have calculated the finite-size spin magnetic susceptibility, $\chi$, which 
is plotted for $N=3$, $N=5$, $N=7$ and $N=9$ in Fig.~\ref{susceptibility}.
The non-interacting result, which in the thermodynamic limit 
approaches $\chi_0 = N (g \mu_B)^2/ (8 k_B T)$ is plotted for 
comparison.  For very high $T$ the Hartree-Fock theory result,
$ \chi^{-1} = 8 k_B (T - \lambda \tilde a(0)/4) / N (g \mu_B)^2 $,
is approached.  However, we see from 
Fig.~\ref{susceptibility} that for $T = T_c^{HF}$, where 
$\chi$ diverges in Hartree-Fock theory, it is in fact enhanced by 
a factor of only $\sim 2.5$ compared to the non-interacting electron
result.  
The inset in this figure expands 
the low-T behavior by plotting $ \ln(\chi) $ vs. $\ln(k_B T)$. 
In the thermodynamic limit this quantity should approach\cite{RS95} 
$ 4 \pi \rho_s/ k_B T $ as $ T \to 0$.  For a finite-size system the 
low $T$ limit of $\chi$ is the Curie susceptibility associated
with the spin-quantum number of the finite-system ground state:
$\chi \to (g \mu_B)^2 S_0 (S_0+1) / 3 k_B T $ so that $ \ln (\chi) $ 
is a linear function of $\ln(T)$ with an offset which increases
with system size.  The low $T$ breaks in the susceptibility plots
indicate that finite-size effects become important for 
temperatures smaller than $\sim 0.1 \lambda/k_B$, roughly consistent
with the temperature below which the microscopic exact diagonalization
and Heisenberg model Monte Carlo calculations differ in 
Fig.~\ref{heisenberg}. 

Comparing all these results we can conclude that while
our self-energy approximation removes the gross failures of
the Hartree-Fock approximation, it still overestimates the 
spin-magnetization at intermediate temperatures.  
Evidently, interactions between spin-waves accelerate the 
decrease of the magnetization with temperature.  We can obtain  
some corroboration of this interpretation by examining the low energy  
portion of the spectrum of the electronic Hamiltonian for $N=13$ shown in
Fig.~\ref{swi}.  These results are for electrons on the 
surface of a sphere.  All excited states with  positive total $L_z$ are
shown.  The largest value of $L_z$ for which a state occurs specifies its
total angular momentum.  The various
eigenenergies are labeled by the occupation numbers of the 
corresponding non-interacting spin-wave states.  Where the 
occupation number exceeds one, the label of the spin-wave
state is repeated so that the number of labels is equal to 
the total spin-wave occupation number of the state.
Single-spin wave states occur in the $S = N/2 -1$ 
portion of the spectrum.  The lowest energy multiple
spin-wave state is the $(1,1)$ state where the $L=1$ 
spin-wave is doubly occupied.  Since there are three degenerate 
spin-waves with $L=1$, if there were no spin-wave
interactions, the six $(1,1)$ two bosons states would be
degenerate and would have an energy equal to twice the 
energy of the $L=1$ single spin-wave state.  
The open horizontal bars in Fig.~\ref{swi} 
indicate the non-interacting spin-wave energies.  
We see there that spin-wave interactions split the 
six $(1,1)$ states into a five fold degenerate $L=2$ level
and a single $L=0$ level.  Generally, multiple spin
wave states are reduced in energy by interactions and 
this effect leads to a magnetization which decreases more
rapidly with increasing temperature than would be expected 
if spin-wave interactions were neglected.  This is most 
apparent in Fig.~\ref{swi} in the effect of 
interactions on the energies of the fifteen expected
$(2,2)$ two-boson states.  When interactions are included
these states are split into $L=4$, $L=2$ and $L=0$ levels. 
We see in Fig.~\ref{swi} that the $L=4$ level, which has
the largest degeneracy, is lowered in energy by
spin-wave interactions. 

We now turn to a comparison between theory and
experiment.\cite{BDPWT95,MAGBPW96}  
An important source of uncertainty is introduced 
by the dependence of the effective interaction between
2D electrons\cite{finitewidth} on the width and, to a lesser 
degree, the height of the quantum well containing the 
electrons.  For a quantum well of width $w$ and infinite barrier 
heights, the effective interaction\cite{AU74} is 
$\tilde V_{eff}(\vec{k}) = F(\vec{k})\tilde{V}(\vec{k})$, where\cite{TBDPW95}  
\begin{equation}
F(k,w)=\frac{32 \pi^{4}(e^{-kw}-1)}{(kw(4\pi^2 + k^2w^2))^2} + \frac{8 \pi^2}{(kw(4\pi^2 + k^2w^2))} 
     + \frac{3kw}{(4\pi^2+k^2w^2)}
\label{form}
\end{equation}
When the system is described by 
phenomenological field-theory or Heisenberg models, the microscopic 
physics enters only through the spin-stiffness.  In
Fig.~\ref{stiffquasi} we plot the spin-stiffness, $\rho_s$, at $\nu =1$ 
as a function of quantum well width for the case of a quantum 
well with infinite barrier heights.  The analytic expression
relating spin-stiffness and effective interaction\cite{GM95} is 
\begin{equation}
\rho_s  = \frac{\lambda \ell^2_c}{32 \pi^2} 
\int_{0} ^{\infty} dk k^3 \tilde V_{eff}(k) \exp ( - k^2
\ell^2_c /2 ).
\label{rhoswidth}
\end{equation} 
The $k^3$ factor in this integrand is responsible for considerable
sensitivity of $\rho_s$ to $\tilde V_{eff}(k)$ at large $k$,
where the finite thickness corrections appear.  For comparison 
the spin-splitting gap $\tilde a(0)$, plotted in Fig.~\ref{stiffquasi},
has a considerably weaker relative dependence on well width.
As shown in Fig.~\ref{stiffquasi} the 
spin-stiffness is reduced by approximately a factor of two
compared to the zero-width 2D layer case for quantum wells
with a width $w = 3 \ell_c$.  This is actually close to the 
typical experimental situation where the quantum well widths
are $\sim 30 nm$ and the fields are $ \sim 10 $ Tesla.
We remark that this reduction of $\rho_s$ reduces the temperature
interval over which collective excitations are dominant and 
the field-theory and Heisenberg phenomenological models are 
appropriate.  In Fig.~\ref{magthick} we compare experimental
data with exact diagonalization calculations of the spin-magnetization
which account for finite-well thickness.  For the purpose of this
comparison, we consider the exact diagonalization results to be 
essentially exact for a model which neglects disorder and 
Landau level mixing.  Compared to the best fits, the experimental
magnetization decreases too slowly at low-temperatures, and 
too quickly at high temperatures.  It seems clear that the experimental
values are too low at the highest temperatures where they fall
below even the non-interacting system magnetizations 
(this happens for $k_B T \sim 0.09$ in Barrett's experiment).  The 
weaker $T$ dependence at low $T$ in the experimental data, could
be due to disorder, and in particular to weak large length scale 
inhomogenity, which is neglected in all theoretical models  
discussed here.

Finally, we comment on the role of skyrmions.
One of the interesting results of the NMR-experiments was the 
experimental evidence for skyrmions at filling factors near one 
\cite{BDPWT95}.  Our diagrammatic theory is not able to account
for skyrmions, although we know that neutral skyrmion-antiskyrmion pair 
excitations exist at $\nu=1$.  Their total energy for $\Delta_z=0$ 
is only half of that of the quasielectron-quasihole pair, 
although the energetic advantage drops quickly 
with increasing Zeeman gap (see  Fig.~1 in
\cite{FBCM94}). Thus, the existence of
such excitations at the upper end of the excitation spectrum should not 
dramatically alter the thermodynamics. 

\subsection{Nuclear spin relaxation rate} 

The optically pumped NMR experiments of Barrett {\it et al.}
can also be used to measure the rate of nuclear spin-relaxation
due to coupling to the electronic spins.  At present, 
measurements of the temperature dependence of the relaxation time
over a broad range of temperatures at $\nu = 1$ are unavailable.
The Korringa theory of nuclear spin relaxation in a 
metal\cite{slichter} can be generalized to electrons in a 
quantum well with the result\cite{berg}
\begin{equation} 
T_1^{-1} = \frac{ k_B T A^2 \Omega^2 |\phi(z)|^4}{(g \mu_B)^2 \hbar} 
 \lim_{\omega \to 0} \frac{{\rm Im} 
\chi^{+-} (\vec r,\vec r; \omega)}{\hbar \omega}.
\label{slichter}
\end{equation} 
In this equation $A$ is the hyperfine coupling constant,
$\Omega$ is the unit cell volume, and $\phi(z)$ is the 
envelope function of the electronic quantum well state.
The influence of interactions on the relaxation rate has been 
studied in the limit where disorder is relatively strong and 
interactions can be treated perturbatively.\cite{vagner,AM91}
If we neglect vertex corrections, the 
local response function $\chi^{+-}$ in Eq.~\ref{slichter}
can be expressed in terms of the spectral function for the 
one-particle Green's function.  The result for the relaxation
rate of nuclei at the center of the quantum well is
\begin{equation} 
T_1^{-1} = C(B,w) \frac{k_B T}{\lambda}
\int_{-\infty}^{\infty} \frac{d E}{(2 \pi)^2} \frac{ - \partial n_F(E)}{\partial E}
\lambda^2 A_{\uparrow}(E) A_{\downarrow}(E). 
\label{rate}
\end{equation}
Inserting GaAs parameters\cite{berg} for the prefactor in 
Eq.~\ref{rate} gives $C(B,w) =  0.47 (B[{\rm T}])^{3/2}/(w[{\rm nm}])^2
{\rm Hz}$.  For typical fields and quantum well widths the prefactor
corresponds to a relaxation time $\sim 100 s$.  The dimensionless
integral in Eq.~\ref{rate}, which gives the relaxation rate in this
unit, is plotted as a function of temperature in Fig.~\ref{relaxrate}.
Existing experimental data cover only the low temperature
limit, and are consistent with the very long relaxation times
indicated here.
Contributions to the relaxation rate come dominantly from the 
continuum portion of the spectral weight near zero energy.
In our theory this is small at both low and high temperatures.
Since $T_1^{-1}$ measures the low-energy spin-flip excitations
of the system, it is amenable to an analysis based on the 
continuum field theory model, which has also been
used to obtain theoretical estimates of its temperature
dependence.\cite{RS95} 
It should, in principle, be possible to extract more information about 
the spectral functions, including information on its behavior far
from the Fermi energy where a microscopic theory is necessary, 
from the 2D-2D tunneling studies of quantum
Hall ferromagnets, which we propose below.  

\subsection{Tunneling Current} 

Electronic spectral functions are traditionally measured 
by tunneling experiments.  The measurement\cite{2d2dtunthry}
of spectral functions 
for 2D electron systems is enabled by techniques\cite{2d2dtunexpt}
for making separate contact between nearby quantum wells.
In the absence of a magnetic field, this technique has 
made it possible to measure the quasiparticle lifetime 
including its dependence on temperature due to carrier-carrier
scattering.  For strong magnetic fields\cite{strfieldthry,strfieldexpt}
the tunneling current is related to the bias voltage by
\begin{equation}
I(V) = \frac{ e t^2 A}{h \lambda \ell_c^2}  \int_{-\infty}^{\infty} \frac{d E}{2 \pi} 
(n_F(E-eV) - n_F(E)) \lambda \sum_{\sigma} A_{\sigma}(E) A_{\sigma}(E-eV) 
\label{tuncurrent}
\end{equation} 
Here, $t$ is 
the tunneling amplitude and $A$ the area of the 2D system.
We caution that the above formula applies when both 2D layers
are kept at filling factor $\nu =1$ in the presence of a bias 
voltage.  In 2D-2D tunneling, layer densities change with 
bias potential because of the finite capacitance of the 
double-layer system, unless a compensating gate voltage 
is applied.  This issue is especially important at 
$\nu =1$ because of the sensitivity of the electron system
to density near this filling factor.
Provided that there is no density change in either layer, 
each spin direction contributes equally to the current. 
Measurements at fractional filling factors have demonstrated 
a deep, wide, and only partially understood minima in the 
spectral function near zero energy.  Our calculations suggest
the possibility of further interesting findings when experiments
are performed at $\nu = 1$.  
In Fig.~\ref{tuniv} we plot the dependence of the 
tunneling current on bias voltage for three temperatures.  
Within our theory a $\delta$-peak with a substantial 
weight proportional to $z z^+ \sim (1-(M/M_0)^2)/4$ appears.  
This peak arises from the product of the two poles 
in the spectral function, and it will occur 
at a temperature dependent 
value of $eV$ equal, for the idealized case of a zero width 
quantum well, at the energy difference 
between the upper and the lower pole positions
plotted in Fig.~\ref{positions}. 
Broader and much weaker features result from the convolution
of a delta-function peak with the continuum continuum contribution, 
and still broader and weaker features from the 
self convolution of the continuum contributions to the spectral 
functions.   Only this last contribution contributes to the linear
tunneling conductance.
In Fig.~\ref{tuniv} we have for visualization purposes arbitrarily
replaced the delta-function contribution by a Lorentzian of 
width $0.01 \lambda$.

We expect that sharp peaks do occur 
at voltages near the exchange splitting,
despite the quantitative limitations of our theory discussed
above.  
In Fig.~\ref{tuncond} we plot
the tunneling conductance,
\begin{equation}
G = \lim_{V \to 0} \frac{I}{V} = 
 \frac{e^2}{h}\frac{t^2 A}{\lambda^2 \ell_c^2}
\int_{-\infty}^{\infty} \frac{d E}{2 \pi} 
\frac{ - \partial n_F(E)}{\partial E} 
\lambda^{2}\sum_{\sigma} A_{\sigma}(E) A_{\sigma}(E)  
\label{tenconduc}
\end{equation} 
as a function of temperature.  The tunneling conductance is 
proportional to the square of the spectral function averaged 
over energy arguments less than $\sim k_B T$.  It has therefore
a temperature dependence similar to that of the 
nuclear spin relaxation rate.  It vanishes at
zero temperature and remains small even for $k_B T$ substantially
in excess of $\Delta_z$ since 
the bulk of the spectral weight is shared between the high
and low-energy poles and exchange enhanced spin splitting 
causes these to be well away from the chemical potential.

\section{Odds and Ends} 

\subsection{Consistency}

A second formally exact expression can be derived which relates
the electronic spectral function to the magnetization.
In the strong field limit, $K=H-\mu N=K_{\uparrow}+K_{\downarrow}$
can\cite{FW71} be written as 
\begin{equation}
K_{\sigma}(T, V, \mu)  =  \frac{N_{\Phi}}{2 \beta}
\sum_{i \nu_{n}} e^{i\nu_n \eta} [i \hbar \nu_n + \xi_{\sigma}^{(0)}] {\cal G}_
{\sigma}(i \nu_n)
\end{equation}
where $\eta$ is a positive infinitesimal and the sum is over Matsubara
frequencies.  Using the spectral representation of the Matsubara Green's
function and performing a contour integral then yields
\begin{eqnarray}
K_{\sigma}(T, V, \mu) & = & \frac{N_{\Phi} }{2} \int_{-\infty} ^{\infty}
 \frac{d E}{2\pi} n_F (E)
 (E + \xi_{\sigma}^{(0)}) A_{\sigma}(E) \; .
\label{K3}
\end{eqnarray}
Using thermodynamic identities, the magnetization can in turn be
expressed in terms of K:
\begin{equation}
M(T,V,\mu) = -\frac{1}{\beta} \int _{0} ^{\beta} d\beta^{'}
\Biggl(\frac{\partial K}{\partial B}\Biggr)_{\scriptstyle \beta^{'},V,\mu }
\label{M2}
\end{equation}
We do not obtain the same result for $M$ from this expression as from 
the more direct expression discussed above if we go beyond the SHF 
approximation. This ambiguity is one 
of several consequences of the fact that our self-energy approximation 
is defined in terms of Hartree-Fock propagators and is not
conserving.\cite{Bay62}

Some partial self-consistency can be achieved by simply 
replacing the SHF occupation factors, $\xi_{\sigma}^{HF}$,   
wherever they appear, by occupation factors calculated 
from the final spectral functions.  At $\nu = 1$, it is sufficient to
specify the difference of majority and minority spin filling
factors, $\Delta \nu$.  Following this procedure 
requires that we solve an equation of the form 
$\Delta \nu = f(\Delta \nu)$, where
$f$ incorporates the entire functional dependence on the r.~h.~s.~of 
Eq.(\ref{M3})).
At the same time, we modify the HF quasiparticle energy accordingly:  
\begin{equation}
\xi_{\uparrow} = - \frac{1}{2}(\Delta_z + \lambda \tilde{a}(0) \Delta \nu) \; .
\end{equation}
Together with Eq.~(\ref{M3}) this procedure defines 
an implicit equation for the difference $\Delta \nu$, {\em i.e.} for the 
magnetization. 
In general this equation has at least one solution with 
$0< \Delta \nu < \Delta \nu^{HF}$. 
We have determined this solution as a function of temperatures
and have found that the solution is unique.
The result is shown in  Fig.~\ref{magn4}. 
 The smaller magnetization 
values are in better accord with experiment.  The improvement probably
does reflect a partial accounting of omissions of the elementary
electron spin-wave scattering theory.  However, the abrupt  
decrease of the magnetization toward non-interacting electron values 
at temperature values $T \sim 0.06 \lambda/k_B$ is certainly unphysical.
Not surprisingly, this {\it ad hoc} procedure does not provide 
satisfying results.  
A more elaborate attempt at a self-consistent scheme, in which the 
full set of equations~(\ref{system}), was solved self-consistently 
has been explored by Haussmann\cite{Hau96} and proved to be equally 
unsatisfying.  

Some hints at possible routes toward a more accurate theory
can be found in examinations of the transverse susceptibility of
a quantum Hall ferromagnet.  
It is remarkable\cite{Kas96} that in the 
LLL, this quantity can be expressed exactly as a geometric series of 
of irreducible particle-hole bubbles ($\bar{\Phi}(\vec{q}, i \omega_n)$)
\begin{equation}
\bar{\chi}^{+-}(\vec{q}, i\omega_n) = 
\frac{\bar{\Phi}(\vec{q}, i \omega_n)}{1+I(\vec{q})\bar{\Phi}(\vec{q},i \omega_n)} \; .
\label{susc_ph}
\end{equation}
where $I(\vec q)$ is independent of frequency.
In the generalized random-phase approximation,
$\bar{\Phi}(\vec{q}, i \omega_n)$ is approximated by 
a bubble with Hartree-Fock propagators.
Generally, the irreducible particle-hole bubble can be expressed
in terms of an irreducible vertex function 
$\bar{\gamma}(\vec{q}; i \nu_n  \uparrow; i(\nu_n
+\omega_n), \downarrow)$. 
\begin{equation}
\bar{\Phi}(\vec{q}, i \omega_n) = 
- (g \mu_B)^2 \frac{e^{-q^2 \ell_c^2/2}}{2\pi \ell_c^2} \frac{1}{\beta} \sum_{i \nu_n} \bar{\gamma}(\vec{q}; i \nu_n \uparrow; i(\nu_n
+\omega_n) \downarrow){\cal G}_{\uparrow}(i \nu_n){\cal G}_{\downarrow}(i(\nu_n + \omega_n))
\label{irrvertex}
\end{equation} 
In the RPA-expression for $\bar{\Phi}$ we have set $\bar{\gamma}=1$
and the GF's are approximated by the SHF-GF's. 
Therefore it is not surprising that any improvement 
of $\cal{G}$ beyond the SHF-GF makes 
a change of $\bar{\gamma}$ necessary in order to satisfy the Goldstone
theorem condition $1=-I(\vec{0})\bar{\Phi}(0,0)$. 

\subsection{Screening}

In any electronic system, screening of electron-electron interactions
plays an important role in the many-particle physics.  
The simplest approximation for the dynamically screened interaction
is the RPA, which, for the present problem,
takes the form: 
\begin{equation}
\tilde{V}(\vec{k}, i \omega_n) = 
\frac{ \tilde{V}_c(\vec{k})}{\epsilon(\vec{k},i \omega_n)} = \frac{\tilde V_c(\vec{k})}{1+\tilde V_c(\vec{k}) \Pi^{0}(\vec{k},
i \omega_n)} = \frac{2 \pi \ell_c}{k + 2\pi \ell_c \Pi^{0}(\vec{k}, i \omega_n)}
\end{equation} 
with the polarization function approximated by 
\begin{equation}
\Pi^{0}(\vec{k}, i \omega_n) = - \frac{e^{-\frac{k^2 \ell_c^2}{2}}}
{2\pi \ell_c^2} \frac{\lambda }{\beta} \sum_{i \nu_n, \sigma} {\cal G}^{HF}_{\sigma}(i \nu_n)
{\cal G}^{HF}_{\sigma}(i(\nu_n - \omega_n)).
\end{equation}
In this approximation, Landau level degeneracy is responsible for 
polarization functions, which vanish at non-zero Matsubara frequencies.
At $i \omega_n = 0$, we find that 
\begin{equation}
\tilde{V}(\vec{k}, i \omega_n =0 ) = 
\frac{\tilde V_c(\vec{k})}
{1+\tilde V_c(\vec{k}) \beta \lambda \; \frac{e^{-\frac{k^2 \ell_c^2}{2}}}{2 \pi \ell_c^2} \sum_{\sigma} n_F(\xi_{\sigma}^{HF})
(1-n_F(\xi_{\sigma}^{HF}))}
\end{equation}
In the long wavelength limit this leads to the following 
expression for the temperature dependent screening wavevector at 
$\nu = 1$ :
\begin{equation}
k_{sc}(T) = \frac{2 \beta \lambda}{\ell_c}  n_F(\xi_{\uparrow}^{HF})(1-n_F(\xi_{\uparrow}^{HF}))
= \frac{\beta \lambda}{2 \ell_c \cosh^2(\beta \xi_{\uparrow}^{HF}(k_{sc},\beta)/2)}
\label{scr3}
\end{equation}
The inclusion of interaction effects in the HF-energies
plays an important role in the temperature dependence of 
$k_{sc}$ implied by this equation.  At low temperatures
the screening wavevector, $k_{sc} \sim \exp( -\xi_{\downarrow}^{HF}/k_B T)/
k_B T$, is extremely small.   
The limited utility of this screening approximation is evidenced by
the discontinuous dependence of $k_{sc}$ on temperature found 
when Eq.~(\ref{scr3}) is solved numerically.

Additional insight into screening in the static long-wavelength limit 
can be obtained by using  Thomas-Fermi theory in which 
\begin{eqnarray}
k_{sc}(T) = 2 \pi \ell_c \lambda n^2 \kappa 
= 2 \pi \ell_c \lambda \biggl( \frac{\partial n}{\partial \mu} \biggr)_{\scriptstyle T,V,N}
=\frac{\lambda}{\ell_c} \biggl( \frac{d \nu}{d \mu} \biggr)_{\scriptstyle T,V,N}\; , 
\label{scr4}
\end{eqnarray}
where $\kappa$ is the compressibility, $n$ is the particle density and $\nu=2\pi \ell_c^2 n$. 
If we neglect the dependence of the spectral function on the
chemical potential $\mu$ 
and exploit particle-hole symmetry at $\nu=1$, we 
obtain
\begin{eqnarray}
k_{sc} = \frac{2 \beta \lambda}{\ell_c} \int_{- \infty}^{\infty} \frac{d E}
	   {(2 \pi)} A_{\uparrow}(E) n_{F}(E)(1-n_{F}(E)).
\label{scr5}
\end{eqnarray}
This is an implicit equation for $k_{sc}$ since 
$A_{\uparrow}(E)$ itself depends on $\xi_{\uparrow}^{HF}$, which again
depends on $\tilde{a}(0)$ and thus on $k_{sc}$ (see (\ref{aofk})). 
With this equation we can use the improved spectral density of
the spin wave-theory 
to estimate the approximate magnitude of wavevector $k_{sc}$ at 
a given temperature. 
Rather than solving self-consistently for $k_{sc}$ at each temperature,
we have fixed $k_{sc}$ at two values and used Eq.~(\ref{scr5}) to 
find those temperatures at which these values are self-consistent. 
As discussed in the previous section, this procedure is adequate 
given the relatively weak sensitivity of our results at low and 
moderate temperatures to the value used for $k_{sc}$. 
At higher temperatures, screening is likely to be important since 
stronger screening causes a flattening of the spin wave
dispersion, which leads in turn to more magnetization suppression,
smaller exchange splittings, more mobile charges, and 
hence still stronger screening.  

\subsection{High Temperature Behavior} 
 
The high temperature expansion of the magnetization gives us some 
insight into the validity
of our approximations and stresses the importance of screening at higher 
temperatures.  However, these results are of rather theoretical 
interest because 
at large temperatures excitations to higher orbital Landau levels become more 
and more probable and our restriction to the LLL becomes questionable. 

In principle, the many-body perturbation theory expansion for the 
thermodynamic potential provides 
a systematic order by order expansion in powers of the 
interaction strength over the bare Zeeman gap, or the temperature
or combinations thereof.
However, the long range of the Coulomb interaction gives
rise to divergent diagrams and complicates issues again; 
for example a logarithmically  divergent contribution   
to the magnetization appears at third order in interaction strength,  
which can be traced back to the divergent second order bubble diagram for the
self-energy.  A systematic way to circumvent this problem is to expand in
terms of interactions, which are screened by 
infinite order bubble diagram partial summations.
Rather than pursue this line, we have attempted to gain some insight from
perturbative expansions by performing the expansion up to
third order in $\beta$ for the case of hard-core interaction model 
with $\tilde V(q)=4 \pi V_0 \ell_c^2\;$ \cite{const}.  

The leading term of the high temperature expansion for the spin magnetization 
is identical to the same term for free particles, {\it i.~e.~}the lowest 
order term in the expansion of  
$M(T)/M_0 = \tanh(\beta \Delta_z/4) = \beta \Delta_z/4 + O(\beta ^2)$. 
As mentioned previously, this limit shows that the Heisenberg 
model, which yields a magnetization that is twice
of this value, fails once itinerancy becomes important.
Our calculations are based on the 
linked-cluster coupling-constant-integration expansion of 
the thermodynamic potential\cite{Kas97}. 
The leading correction, which is quadratic in $\beta$,  
is determined solely by the exchange integral $2 V_0$ and
leads to an increase of the spin magnetization. 
We have carried this expansion out to third order and find that 
\begin{equation}
M = M_0 \frac{\beta \Delta_z}{4} \Bigl \lbrace 1 + \frac{\beta V_0 \lambda}{2} 
+ \frac{\beta^2}{8} ( V_0^2 \lambda^2 - \frac{\Delta_z^2}{6} ) \Bigr \rbrace 
+ O (\beta^4)
\label{v0}
\end{equation}
Note that the $V_0^2$ term also contributes positively to the
magnetization.
Fig.~\ref{hight3} shows $M T /M_0$ vs.~$1/T$ at high temperatures.
We can see from this figure that beyond leading order interaction
contributions become important for $k_B T$ smaller than
$\approx 2.5 \lambda$. For comparison the results from exact diagonalizations 
for fixed particle numbers are given, which show the right slope but 
exhibit finite size corrections of the expected order of magnitude \cite{finsiz}.  
 Obviously the importance of correlations
in this system persists to high temperatures.  This 
property is a key in validating the use of models,  
which correctly describe only collective fluctuations
in magnetization, even when $k_B T $ is not small
compared to underlying interaction energy scales. 
 
\section{Summary and Conclusions}

In this paper we report on a study of the one-particle
Green's function of a quantum Hall system at filling factor $\nu=1$.
The ground state of the two-dimensional electron 
system in this case is a strong ferromagnet.
Many analogies exist between the properties and the 
theoretical description of this system and conventional
metallic ferromagnets.  At $\nu=1$ the ground state and 
the elementary excitations of quantum Hall ferromagnets
are given exactly by time-dependent Hartree-Fock theory.
This success is analogous to the success of band theory 
in describing the ground state and both collective and 
particle-hole elementary excitations of band ferromagnets.
At finite temperatures, however, we show that Hartree-Fock
theory fails qualitatively for quantum Hall ferromagnets,
just as band theory fails for metallic ferromagnets.  
Our work is based on an improved approximation for the 
electron self-energy, which describes the scattering of 
fermionic quasiparticles off the spin-wave collective 
excitations composed of coherent combinations of 
spin-flip particle-hole excitations.  This perturbative approximation
is equivalent to ones, which have been used\cite{HE73} for 
models of itinerant electron ferromagnets at finite
temperatures.  Here we have the advantage that complicated
band-structures do not confuse the comparison of theory and 
experiment.  We find that at intermediate temperatures
where the density of spin-wave excitations is high, although 
our approximation gives a huge improvement over Hartree-Fock
theory, it still overestimates the magnetization by nearly
a factor of two.  We attribute this failure to the neglect
of interactions between spin-waves in our approximation.  
Nevertheless, we expect that the qualitative physics predicted by our
approximation is correct.
On the basis of our calculations we predict that a 
sharp peak will occur in 2D-2D tunneling current when the 
bias energy $eV$ is approximately equal to the spin-splitting
and that the strength of this peak will be approximately 
proportional to temperature at low $T$.  This peak is due
to fluctuations in the direction of the exchange field, which
separates the energy of majority and minority spin quasiparticles.
We have recently argued\cite{prljmr} that in 
metallic ferromagnets, this mechanism is responsible for 
the temperature dependence of magnetoresistance in ferromagnetic
tunnel junctions.  In that case, however, the mechanism 
cannot be directly verified by tunneling experiments because
the width of the quasiparticle bands is comparable to their 
exchange-spin-splitting.  Verification of the predicted 
effect in quantum Hall ferromagnets, therefore, has important
implications for metallic magnetic tunnel junctions.

\section{Acknowledgements}

We thank S.~Barrett for sharing data from the NMR-measurements of 
his group prior to publication.
Discussions with W.~Apel, S.~M.~Girvin, B.~Goldberg,
C.~Hanna, R.~Haussmann, H.~Mori, and S.~Sachdev  
are gratefully acknowledged.
One of the authors (M.~K.)~was supported by a fellowship of the
German Academic Exchange Service (DAAD).
This work was supported in part by NSF grant No.~DMR97-14055.

\appendix
\section{Feynman rules}
\label{feynman} 

\begin{figure}
\vskip 4.0cm
\centering
\epsfxsize=5.0 cm
\epsfysize=3.0 cm
\leavevmode
\epsffile{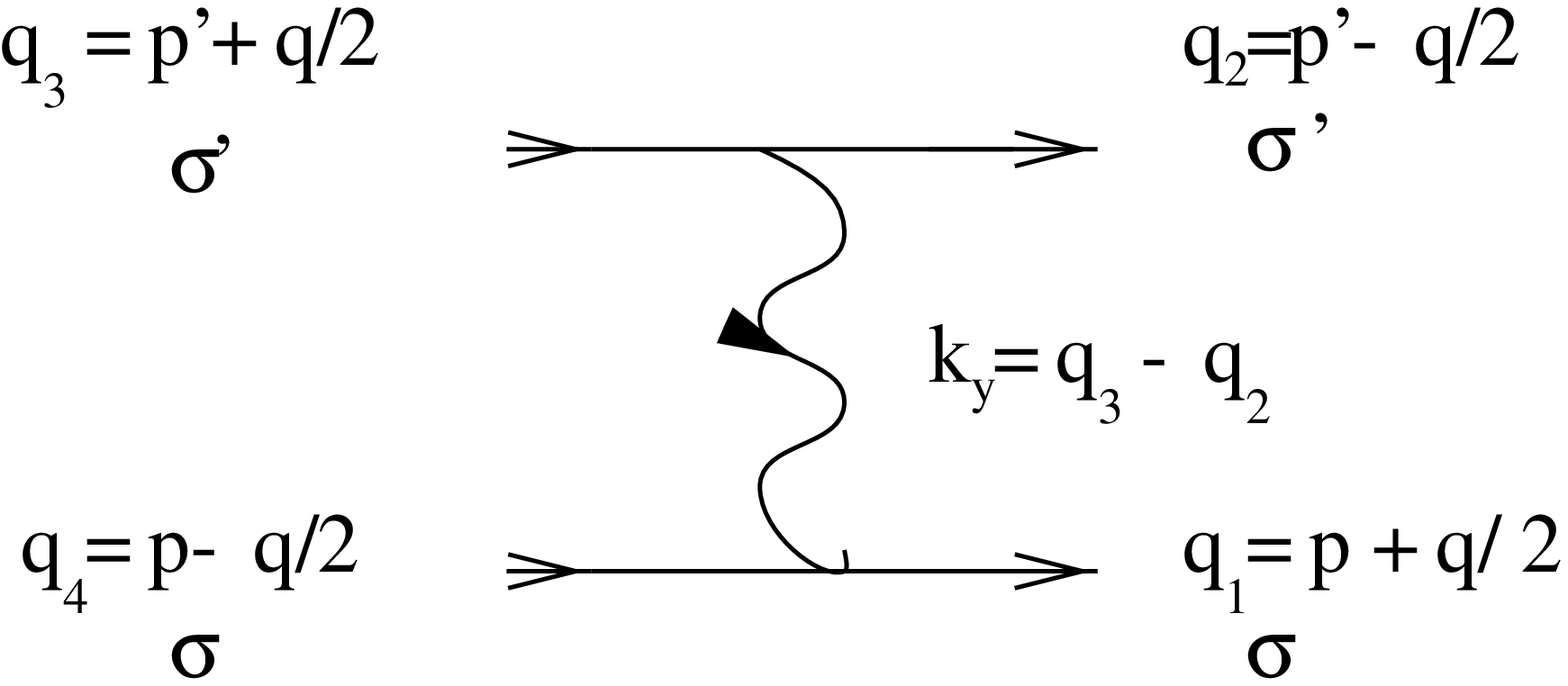}
\caption[]{ Landau gauge electron-electron interaction vertex.}
\label{vert}
\end{figure}

\begin{figure}
\vskip 1.0cm
\centering
\epsfxsize=10.0 cm
\epsfysize=8.0 cm
\leavevmode
\epsffile{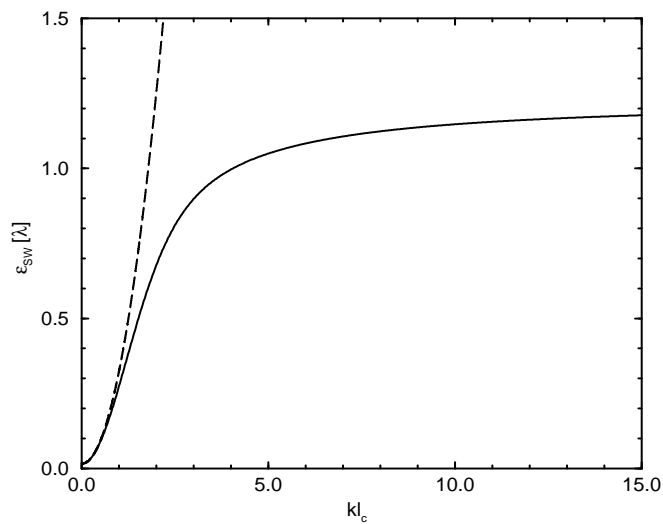}
\caption[]{
Spin-wave dispersion $\epsilon_{SW}(k)$ at
zero temperature with a Zeeman gap $\Delta_z = 0.016 \lambda$ and a static
screening wavevector $k_{sc}=0.01 l_{c}^{-1}$.  Note the small
value of $\Delta_z $ compared with the spin-wave
bandwidth $\lambda \tilde{a}(0)$.  The dashed line shows 
$\Delta_z + 4 \pi \rho_s k^2 \ell_c^2$.  
} 
\vskip 1cm
\label{disp}
\end{figure}

\begin{figure}
\centering
\leavevmode
\epsffile{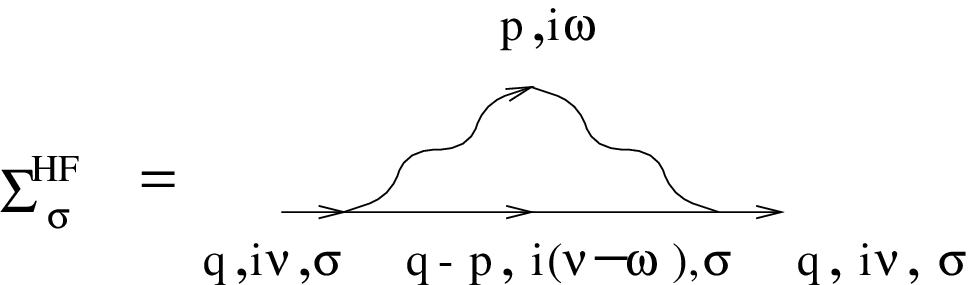}
\vskip 0.5cm
\caption[]{
Proper self-energy diagram in the self-consistent HF approximation.
The propagator line in this diagram must be determined 
self-consistently.  This approximation leads to a frequency
independent self-energy and hence to a Green's function whose spectral
weight consists of a single $\delta-$function.
The first order tadpole 
diagram is absent because of the introduction of a neutralizing 
background charge.} 
\label{diag}
\end{figure}

\begin{figure}
\vskip 1.0cm
\centering
\epsfxsize=10.0 cm
\epsfysize=8.0 cm
\leavevmode
\epsffile{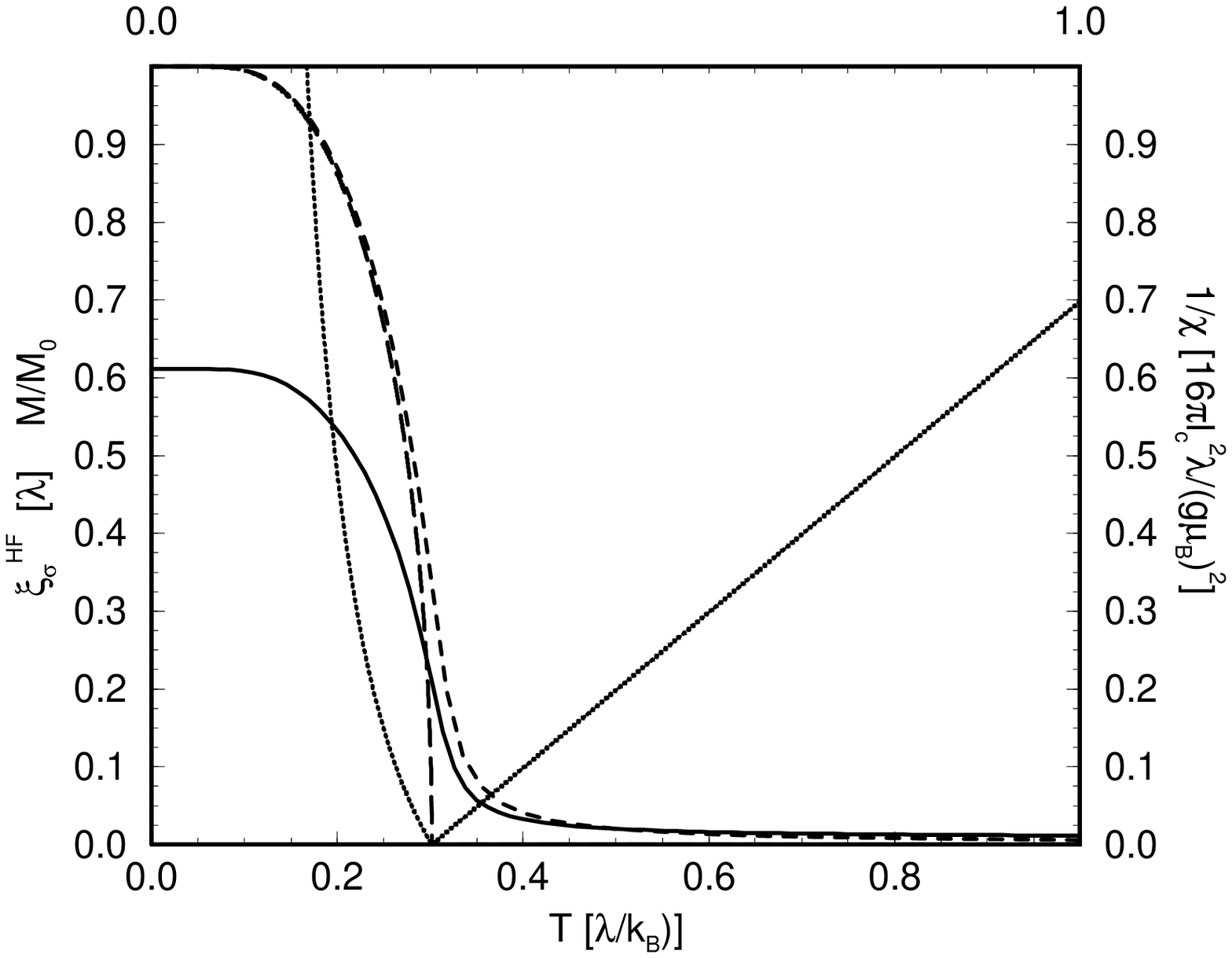}
\caption[]{Hartree-Fock eigenenergy $\xi_{\downarrow}^{HF}$ 
(solid line) as a function of temperature at $\nu =1$;  
$\xi_{\uparrow}^{HF}= -\xi_{\downarrow}^{HF}$ because of particle-hole 
symmetry at $\nu=1$. The magnetization $M=M_0 
(\nu_{\uparrow}-\nu_{\downarrow})$ 
within the SHF is depicted for 
$\Delta_z=0.016 \lambda$ (dashed curve) and $\Delta_z=0.0 \lambda$ 
(long dashed curve), 
respectively. 
Note the finite magnetization at low $T$ in the latter case 
incorrectly indicating the existence of an ordered phase for $T$ below 
$T_c=\tilde{a}(0) \lambda/(4 k_B)$. The uniform static inverse susceptibility 
is plotted as a dotted line in units of $(16 \pi l_c^2 \lambda)/(g \mu_B)^2$ for 
$T \ge T_c$ as well for $T < T_c$.
}
\vskip 1cm
\label{schf}
\end{figure}

\begin{figure}
\vskip 0.5cm
\centering
\epsfxsize=16.0 cm
\epsfysize=3.5 cm
\leavevmode
\epsffile{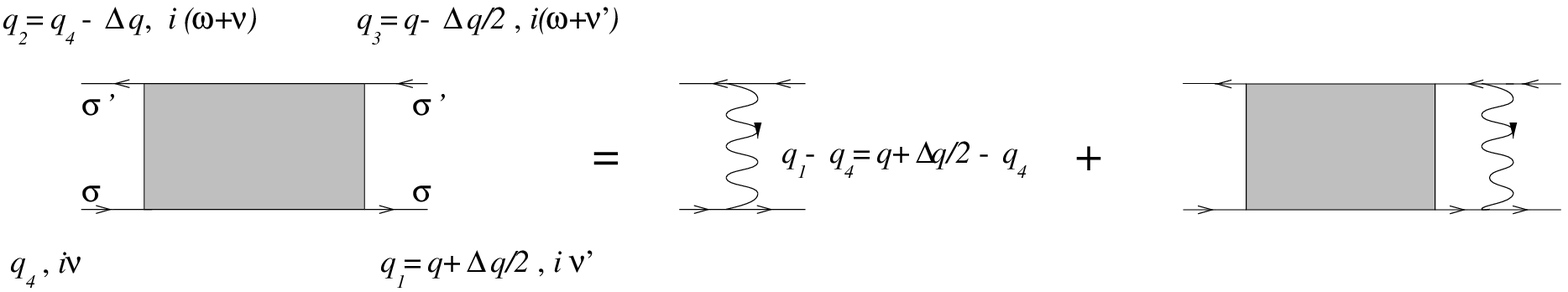}
\caption[]{
The self-consistent integral equation for the scattering vertex $\Gamma^{(4)}$ 
in the particle-hole ladder approximation. 
}
\vskip 1cm
\label{leit}
\end{figure}

\begin{figure}
\vskip 1.0cm
\centering
\epsfxsize=10.0 cm
\epsfysize=8.0 cm
\leavevmode
\epsffile{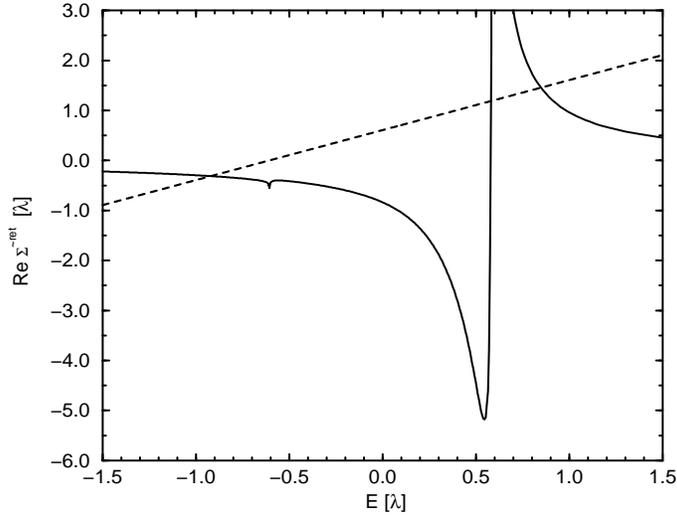}
\caption[]{ 
The real part of the self-energy $\tilde{\Sigma}_{\uparrow}^{ret}(E)$
and the line $E - \xi_{\uparrow}^{HF}$ as a function of 
$E$ at temperature $T=0.1 \lambda/k_B$ ($\Delta_z=0.016 \lambda, 
k_{sc}=0.01 \ell_c^{-1}, w=0.0 \ell_c$). 
The quasiparticle poles occur at the two energy values 
$E^{-}_{\uparrow}$ and
$E^{+}_{\uparrow}$, at which the curves intersect.
Because the self-energy diverges in opposite directions as 
the upper and lower boundaries of the 
central branch-cut interval, two poles exist at any
temperature in this approximation. 
}
\vskip 1cm
\label{real}
\end{figure}

\begin{figure}
\vskip 1.0cm
\centering
\epsfxsize=10.0 cm
\epsfysize=8.0 cm
\leavevmode
\epsffile{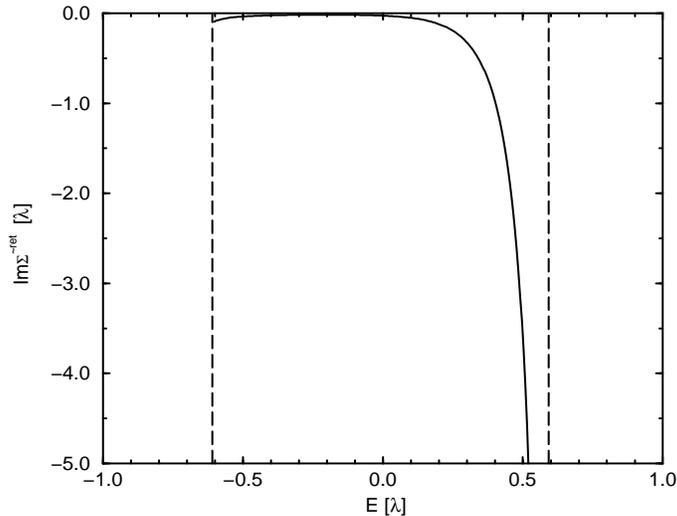}
\caption[]{
The imaginary part of the spin up self-energy 
$\tilde{\Sigma}_{\uparrow}^{ret}(E)$ at $T=0.1 \lambda/k_B$.
This quantity is non-zero only, {\em i.e.} within the central
interval $I_{\uparrow}$ defined in the text.  
Note that the 
imaginary part tends to a finite value on the right boundary of 
this interval. The parameters are the same as in Fig.~\ref{real}. 
}
\vskip 1cm
\label{imag}
\end{figure}

\begin{figure}
\vskip 1.0cm
\centering
\epsfxsize=10.0 cm
\epsfysize=8.0 cm
\leavevmode
\epsffile{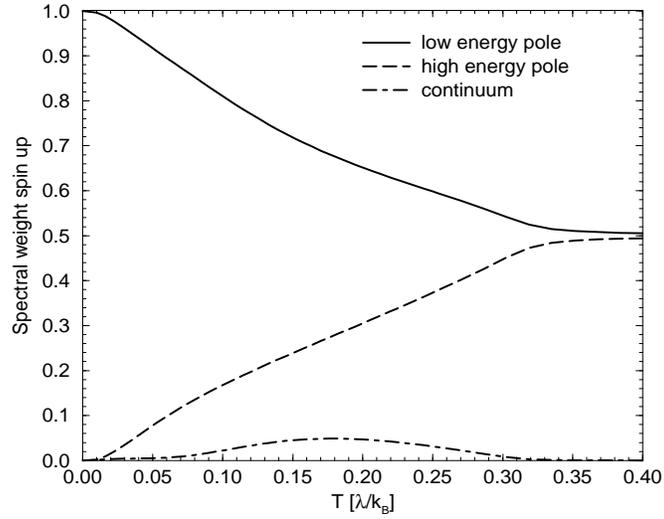}
\caption[]{
Temperature dependence of the partitioning of the majority spin
Green's function spectral weight between
low and  high energy poles and the intermediate energy continuum.
}
\vskip 1cm
\label{weights}
\end{figure}

\begin{figure}
\vskip 1.0cm
\centering
\epsfxsize=10.0 cm
\epsfysize=8.0 cm
\leavevmode
\epsffile{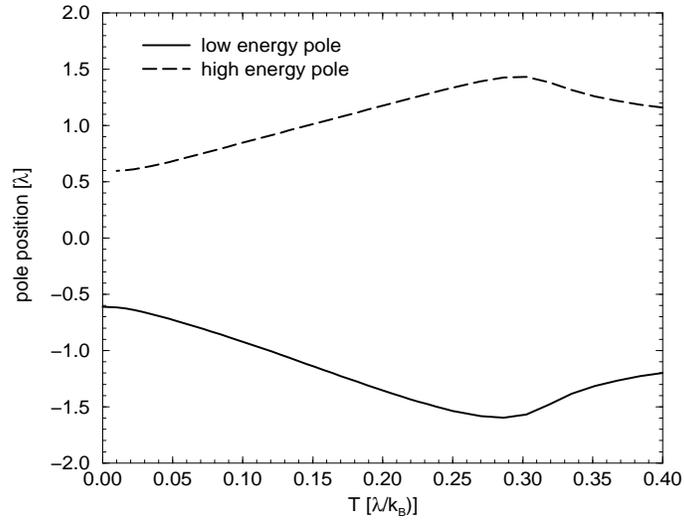}
\caption[]{
Positions of the low and high energy poles in the majority 
spin Green's function $G^{ret}_{\uparrow}(E)$ as a 
function of temperature ($\Delta_z=0.016 \lambda, k_{sc}=0.01 \ell_c^{-1},
w=0.0$).
}
\vskip 1cm
\label{positions}
\end{figure}

\begin{figure}
\centering
\epsfxsize=10.0 cm
\epsfysize=8.0 cm
\leavevmode
\epsffile{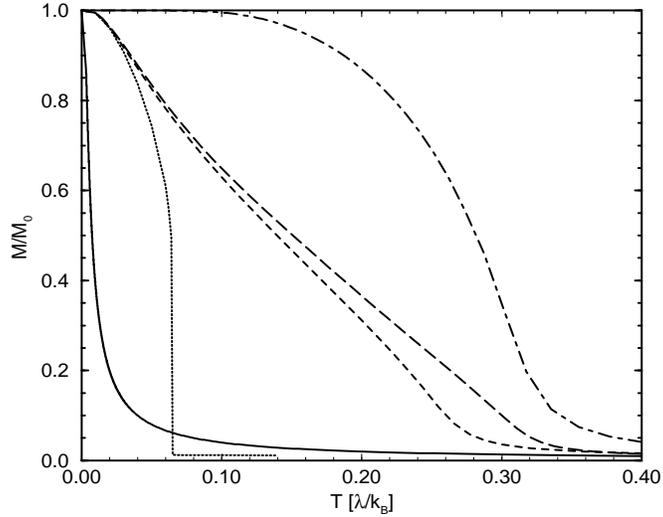}
\caption[]{
Results for $M(T)$ for $\Delta_z=0.016 \lambda $ and zero width 
of the quantum well. The solid curve is the spin magnetization 
for free electrons with the bare Zeeman energy $\Delta_z$. 
For comparison, the dot-dashed curve is the result
of the SHF ($k_{sc}=0.01 l_c^{-1}$) theory. 
The long-dashed and short-dashed lines show the results of the
present theory with two different choices for the {\it ad hoc}
screening vector, $k_{sc}=0.01 l_c^{-1}$ and $k_{sc} = 0.1 l_c^{-1}$, 
which correspond to temperatures $0.09 \lambda/k_B$ and $0.18 \lambda/k_B$, 
respectively. 
The magnetization results for a partially selfconsistent theory, which 
is discussed in Section VI, are shown as dotted curve ($k_{sc}=0.01 \ell_c^{-1}$). 
}
\vskip 1cm
\label{magn4}
\end{figure}

\begin{figure}
\vskip 1.0cm
\centering
\epsfxsize=10.0 cm
\epsfysize=8.0 cm
\leavevmode
\epsffile{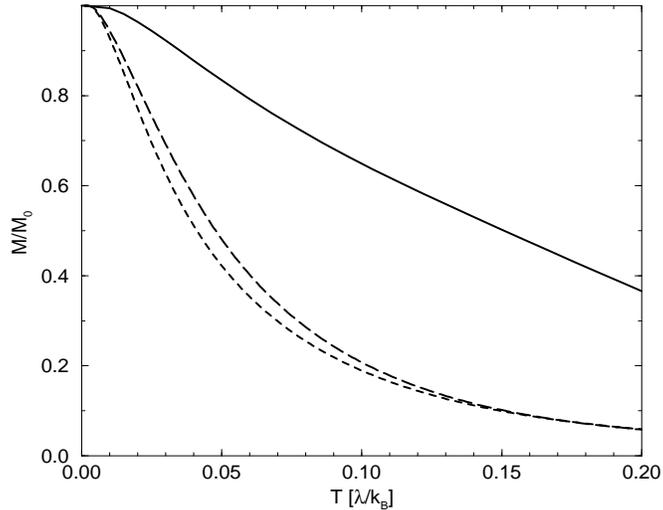}
\caption[]{
Results for the magnetization from finite size
diagonalizations ($N=9$ - long-dashed line, $N=7$ - dashed line)
in the spherical 
geometry for the ideal, unscreened Coulomb interaction.  
This is compared with the result from our perturbative 
approximation for the self-energy (solid curve) where 
we assumed the screening wavevector $k_{sc}=0.01 l_c^{-1}$. 
The quantum well has zero width and the Zeeman energy is 
$\Delta_z=0.016 \lambda$.
}
\label{finsize}
\end{figure}

\begin{figure}
\centering
\epsfxsize=10.0 cm
\epsfysize=8.0 cm
\leavevmode
\epsffile{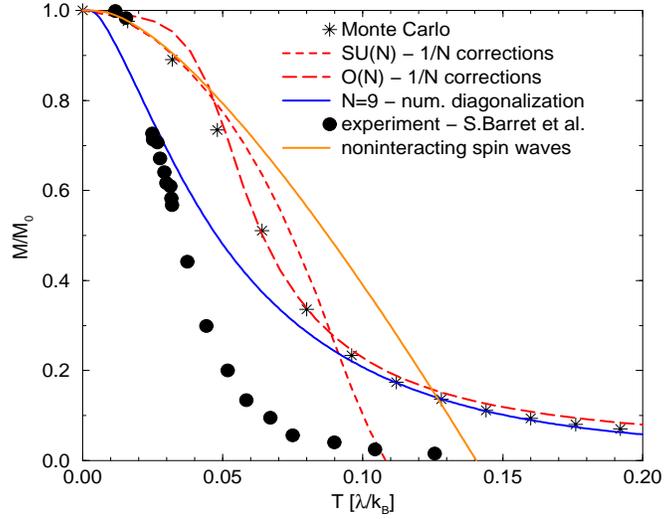}
\caption[]{
Results for the continuous field theory: the dots show 
results of a Monte Carlo calculation. The other curves 
refer to the mean field theories for SU($N$) and O($N$) models 
including $1/N$-corrections as described in Ref.~\cite{timm}.  
For comparison the results from the exact diagonalization for nine particles 
on a sphere, the experimental data of Barrett et al. and 
a naive calculation assuming non-interacting spin-waves 
with a quadratic dispersion relation are added. 
}
\vskip 1cm
\label{heisenberg}
\end{figure}

\begin{figure}
\centering
\epsfxsize=10.0 cm
\epsfysize=8.0 cm
\leavevmode
\epsffile{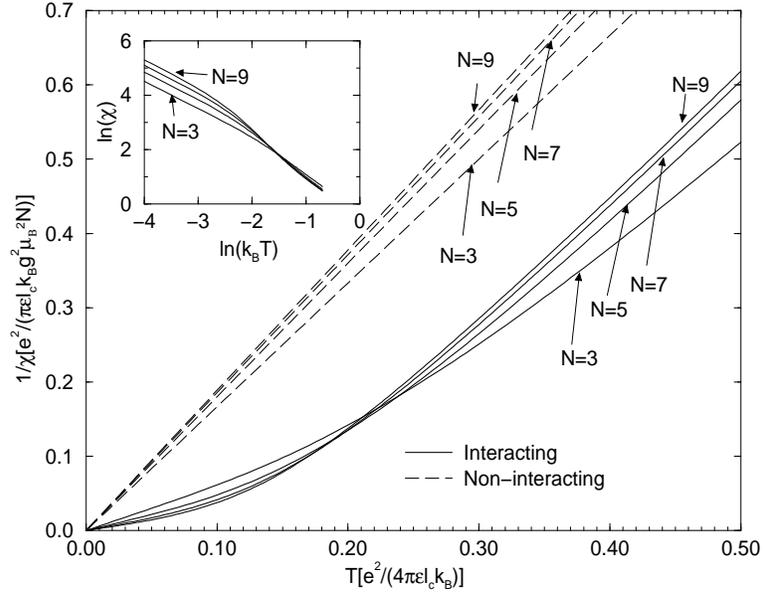}
\caption[]{
Numerical finite-size results for the inverse magnetic susceptibility
in units of $ 4 \lambda/ ((g \mu_B)^2 N$.  In these units the 
free electron result is $\chi^{-1}=2 (k_B T/\lambda)$ and 
the Hartree-Fock result is $\chi^{-1} =  2 
(k_B T/ \lambda - 0.3133)$.
}
\vskip 1cm
\label{susceptibility}
\end{figure}

\begin{figure}
\centering
\epsfxsize=10.0 cm
\epsfysize=8.0 cm
\leavevmode
\epsffile{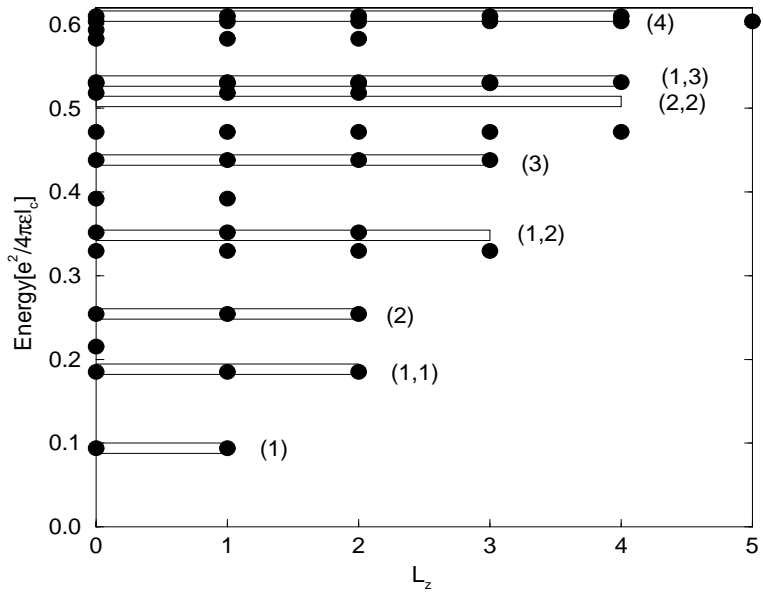}
\caption[]{
Low energy positive $L_z$ excitations for thirteen electrons 
on a the surface of a sphere at $\nu =1$.
The labels are assignments of linear spin-wave occupations corresponding
to the exact eigenstates.
}
\vskip 1cm
\label{swi}
\end{figure}

\begin{figure}
\centering
\epsfxsize=10.0cm
\epsfysize=8.0cm
\leavevmode
\epsffile{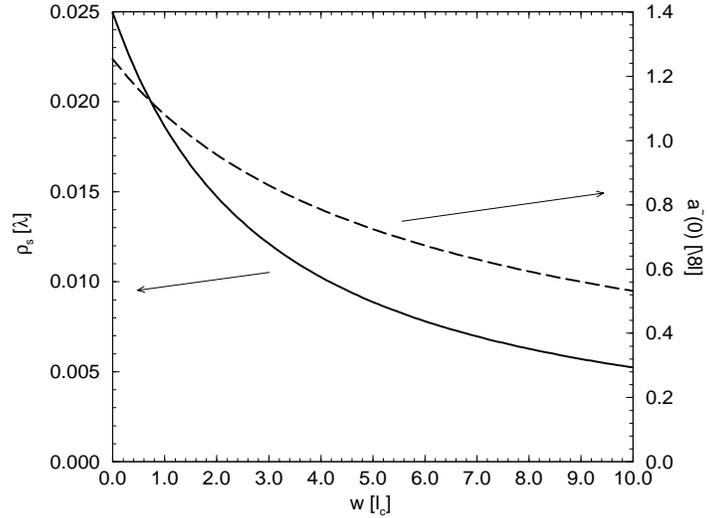}
\caption[]{
Spin-stiffness $\rho_s$ (solid line) and 
spin-splitting gap $\tilde a(0)$ (long-dashed curve) for an infinite height 
quantum well as a function of well width $w$ assuming an ideal Coulomb 
interaction. These results become exact in the strong field limit
where Landau level mixing can be neglected.  The scale for the 
spin-splitting energy is on the right and that for the spin-stiffness
is on the left.
}
\vskip 1cm
\label{stiffquasi}
\end{figure}

\begin{figure}
\centering
\epsfxsize=10.0 cm
\epsfysize=8.0 cm
\leavevmode
\epsffile{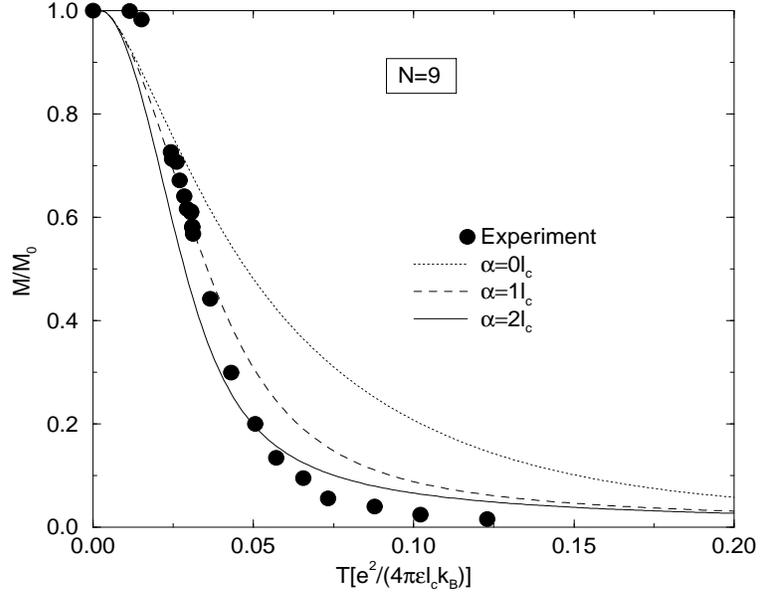}
\caption[]{
Temperature dependence of the spin magnetization for nine particles 
as a function of temperature for quantum wells of different
thicknesses compared with experiment \cite{BDPWT95}.  
In this case the quantity $\alpha$ describes the width 
of a Gaussian charge distribution instead of the 
hard-wall quantum well used elsewhere.
The weak temperature 
dependence at low temperatures in experiment
may be due to disorder.
}
\vskip 1cm
\label{magthick}
\end{figure}

\begin{figure}
\centering
\epsfxsize=10.0 cm
\epsfysize=8.0 cm
\leavevmode
\epsffile{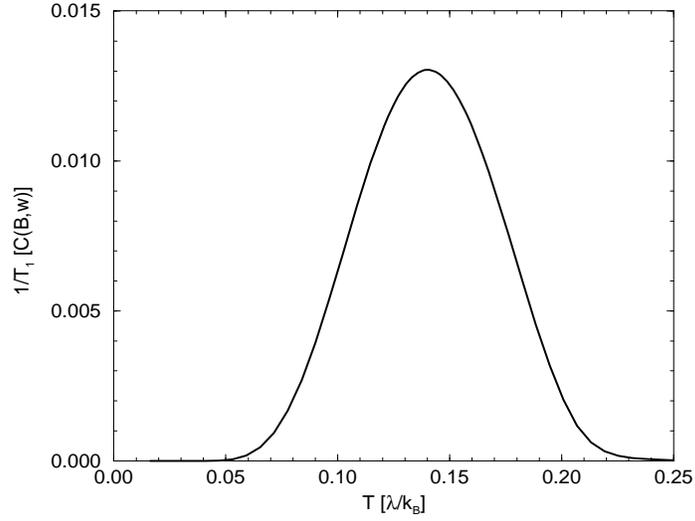}
\caption[]{
Nuclear spin relaxation rate as a function of temperature in
units of the field and quantum well width dependent prefactor $C(B,w)$ 
discussed in the text. The prefactor becomes $9.8\; 10^{-3} Hz$
for a magnetic field $B=7.05 T$ and 
$w=3.11 \ell_c$ ($k_{sc}=0.01 \ell_c^{-1}$). } 
\vskip 1cm
\label{relaxrate}
\end{figure}

\begin{figure}
\centering
\epsfxsize=10.0 cm
\epsfysize=8.0 cm
\leavevmode
\epsffile{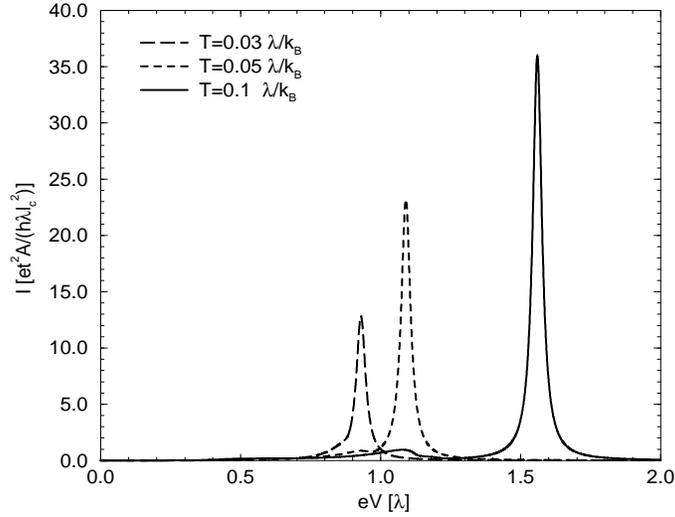}
\caption[]{
Tunneling current-voltage relation as a function of bias 
voltage at $\nu=1$ for three temperatures 
$T = 0.03, 0.05$ and $0.1 \lambda/k_B$ where the $\delta$-peaks of 
the spectral functions are replaced by Lorentzians with width 
$\epsilon=0.01 \lambda$. The 
parameters used for this calculation are: 
$\Delta_z=0.016 \lambda, w=3.11 \ell_c, k_{sc}=0.01 \ell_c^{-1}$. 
}
\label{tuniv}
\end{figure}

\begin{figure}
\centering
\epsfxsize=10.0 cm
\epsfysize=8.0 cm
\epsffile{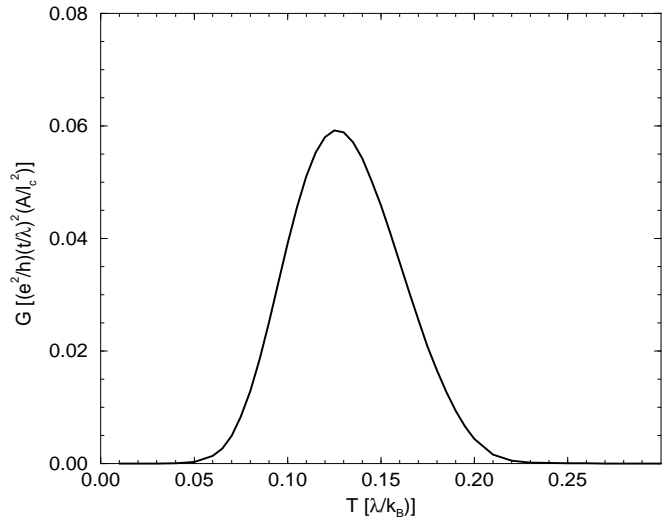}
\caption[]{
Tunneling conductance in the limit of vanishing voltage 
as a function of temperature at $\nu=1$ for the quantum well 
width $w=3.11 \ell_c$ and $\Delta_z=0.016 \lambda$ 
($k_{sc}=0.01 \ell_c^{-1}$). 
}
\label{tuncond}
\end{figure}

\begin{figure}
\centering
\epsfxsize=10.0 cm
\epsfysize=8.0 cm
\leavevmode
\epsffile{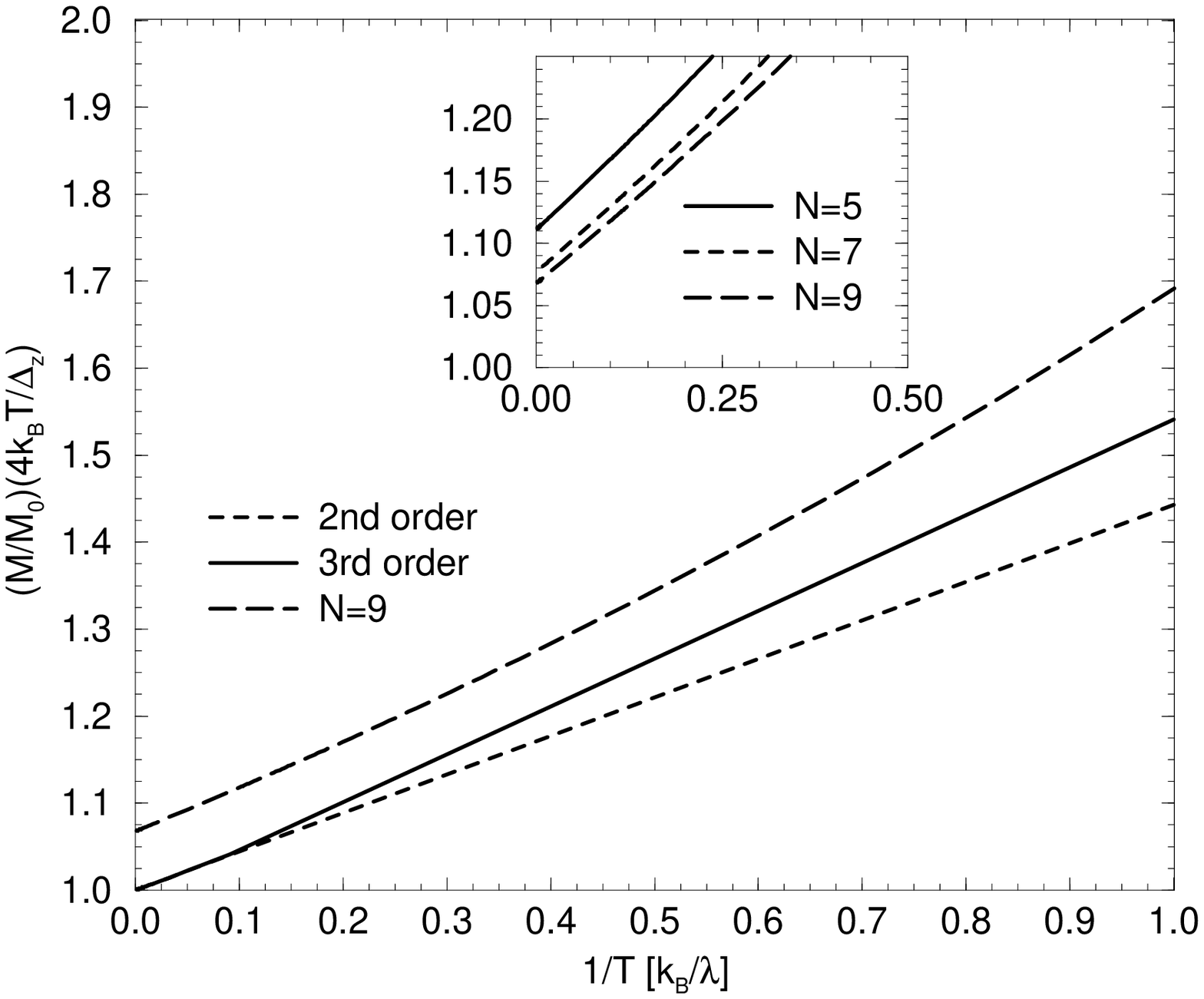}
\caption[]{
Spin magnetization to third order in $1/T$ normalized by the 
high-temperature free particle magnetization ($\Delta_z/
(4 k_B T)$) for the hard-core model 
($V_0=\sqrt{\pi}/2,\Delta_z=0.016 \lambda, w=0.0$).
The resulting deviation of the nine particle exact diagonalization is 
due to finite size effects whose diminishing influence with increasing 
particle number is shown 
in the inset. 
}
\vskip 1cm
\label{hight3}
\end{figure}

We list here the Feynman rules for the Landau gauge in the LLL 
we used to evaluate 
self-energy diagrams.  For a self-energy diagram 
of $n$th order with $(2n-1)$ 
internal electron lines the rules are  
\cite{AB91}: 
\begin{enumerate}
\item Label all directed lines with $y$--momentum $k_y$, Matsubara frequency
$i \nu_{n}$ and spin index $\sigma$ as well as each of the $n$ directed 
interaction lines with a two-dimensional wavevector $\vec{k}$. 
Conserve the $y$--component of the momentum and the spin at each vertex. 
\item
Assign an unperturbed GF ${\cal G}^{(0)}_{\sigma}(i\nu_{n})$ to each 
line as well as a factor $\sum_{\vec{k}}
\tilde{V}(\vec{k})e^{-k^2\ell_c^2/2} e^{ik_x(q_1-q_3) \ell_c^2}
\delta_{k_y,q_3-q_2}$ to each vertex. Here, $q_1$ stems from the outgoing 
line of one particle and
$q_3$ from the ingoing line of the other one (see Fig. \ref{vert}). 
\item
Sum up over all $n$ free internal $k_y$ of the electron lines, 
the independent 
Matsubara frequencies and spin indices. 
\item
Multiply the expression by the factor 
$(\frac{-1}{\beta})^{n}A^{-n} (-1)^{F}$ 
where $F$ is the number of closed fermionic loops. 
\item
At closed fermionic loops  summations  of the type 
\begin{equation}
\frac{1}{A} \sum_{p} e^{ip
(k_{1,x}+ \ldots + k_{m,x})\ell_c^2} = \frac{N_{\Phi}}{A} \delta_{\; \sum_{i=1}^{m} k_{i,x};0}
= \frac{1}{2\pi \ell_c^2} \delta_{\; \sum_{i=1}^{m} k_{i,x};0}
\end{equation}
occur guaranteeing the conservation of the transferred momenta. 
\end{enumerate}

\noindent

\noindent

\noindent

\noindent

\noindent

\noindent

\noindent

\noindent

\noindent

\noindent

\noindent

\noindent

\noindent

\begin{thebibliography}{99}

\bibitem{MFB96}
A.~H.~MacDonald, H.~A.~Fertig,
L.~Brey, Phys.~Rev.~Lett.~{\bf 76}, 2153 (1996). 

\bibitem{GM96}
For a review of quantum Hall ferromagnets see S.~M.~Girvin
and A.~H.~MacDonald 
in {\em Novel Quantum liquids in Low-Dimensional Semiconductor 
Structures}, S.~Das Sarma and A.~Pinczuk (eds.), (Wiley, New York, 1996).
For a brief and elementary discussion see A.~H.~MacDonald, 
Solid State Commun. {\bf 102}, 143 (1997).

\bibitem{moriya} For reviews of theoretical work on itinerant
electron ferromagnets see C. Herring in {\it Magnetism}, Volume 4,
edited by G.~T.~Rado and H.~Suhl, (Academic, New York, 1966) and T.~Moriya,
{\it Spin Fluctuations in Itinerant Electron Magnetism} (Springer--Verlag,
Solid State Science, Vol.~56,
Berlin, 1985).

\bibitem{BDPWT95}
S.~E.~Barrett, G.~Dabbagh, L.~N.~Pfeiffer, K.~W.~West, and R.~Tycko,
Phys.~Rev.~Lett.~ {\bf 74}, 5112 (1995).

\bibitem{MAGBPW96}
M.~J.~Manfra, E.~H.~Aifer, B.~B.~Goldberg, D.~A.~Broido, L.~Pfeiffer, 
and K.~West, Phys.~Rev.~B~{\bf 54}, R17327 (1996).

\bibitem{RS95}
N.~Read and S.~Sachdev, Phys.~Rev.~Lett.~{\bf 75}, 3509 (1995). 

\bibitem{timm}
C.~Timm, S.~M.Girvin, P.~Henelius, and A.~W.~Sandvik, 
Phys.~Rev.~B~{\bf 58}, 1464 (1998).

\bibitem{remarkrhos}
As we see below, however, the present work suggests that in 
the case of long-range electron-electron interactions 
the spin-stiffness energy may {\it not} completely 
determine the low-temperature, weak Zeeman coupling 
physics.  

\bibitem{KM96}
Marcus Kasner, A.~H.~MacDonald, Phys.~Rev.~Lett.~{\bf 76}, 3204 (1996).

\bibitem{Hau96}
R.~Haussmann, Phys.~Rev.~B  {\bf 53}, 7357 (1996); 
Phys.~Rev.~ B  {\bf 56}, 9684 (1997). 

\bibitem{JG89} 
R.~O.~Jones, O.~Gunnarson, Rev.~Mod.~Phys. {\bf 61}, 689 (1989).

\bibitem{HF84} See, for example, A.H. MacDonald, 
Phys. Rev. B {\bf 30}, 4392 (1984). 

\bibitem{remarkhf} The undependable nature of conclusions 
based on pure Hartree-Fock or Hund's rule arguments is
emphasized by their failure away from $\nu = 1$. 

\bibitem{polar} This simple result for $\nu=1$ can not immediately
generalized to other QHE filling 
factors. {\em E.g.}, at filling factor $\nu=2/3$ the ground state 
appears to be a spin singlet for $g=0$.
See for example T.~Chakraborty and P. Pietil\"ainen, Phys.~Rev.~Lett.~
{\bf 76}, 4018 (1996) and work cited therein. 

\bibitem{GM95}
Kun Yang, K.~Moon, L.~Zheng, A.~H.~MacDonald, S.~M.~Girvin, D.~Yoshioka,
S.-C.~Zhang, Phys.~Rev.~Lett.~ {\bf 72}, 732 (1994);
K.~Moon, H.~Mori, Kun Yang, S.~M.~Girvin, A.~H.~MacDonald, L.~Zheng,
D.~Yoshioka, S.-C.~Zhang, Phys.~Rev.~B {\bf 51},  5138 (1995).

\bibitem{Kas96}
M.~Kasner, unpublished.

\bibitem{BIE81}
Yu.~A.~Bychkov, S.~V.~Iordanskii, G.~M.~Eliashberg,
Pis'ma \v{Z}.~Eksp.~Teor.~Fiz.~{\bf 33}, 152 (1981),
(JETP Letters {\bf 33}, 143 (1981)).

\bibitem{KH84}
C.~Kallin, B.~I.~Halperin, Phys.~ Rev.~B {\bf 30}, 5655 (1984).

\bibitem{RM86}
M.~Rasolt, A.~H.~MacDonald, Phys.~Rev.~B {\bf 34}, 5530 (1986);
A.H. MacDonald and G.C. Aers, Phys. Rev. B {\bf 34}, 2906 (1986).

\bibitem{Mac84} A.H. MacDonald, J. Phys. C {\bf 18}, 1003 (1985).

\bibitem{limit}
In the case of no screening $k_{sc}=0$, we recover the well-known 
explicit expression $\tilde{a}(k) = 
\sqrt{\pi/2}e^{-k^2 \ell_c^2/4}I_{0}(k^2 \ell_c^2/4)$. 

\bibitem{leshouches} See for example, 
A.H. MacDonald, in {\it Proceedings of the  Les Houches Summer School on
Mesoscopic Physics} (North-Holland, Amsterdam, 1995), edited by  E. Akkermans,
G. Montambeaux, J.-L. Pichard, and J. Zinn-Justin.

\bibitem{skyrmion} S. L. Sondhi, A. Karlhede, S. A. Kivelson, and E. H.
Rezayi, Phys. Rev. B {\bf 47}, 16419 (1993);
H.A. Fertig, L. Brey, R. Cote, and A.H. MacDonald, 
Phys. Rev. B {\bf 50}, 11018 (1994); 
A. H. MacDonald, H.  A. Fertig, and L. Brey,
Phys. Rev. Lett. {\bf 76}, 2153 (1996);
H.A. Fertig, Luis Brey, R. C\^{o}t\'{e}, A.H. MacDonald,
A. Karlhede, and S.L. Sondhi,
Phys. Rev. B {\bf 55}, 10671 (1997); 
M. Abolfath, J.J. Palacios, H.A. Fertig,
S.M. Girvin, and A.H. MacDonald, Phys. Rev. B {\bf 56}, 6795 (1997).

\bibitem{FW71}
A.~L.~Fetter, J.~D.~Walecka, {\it Quantum Theory of Many-Particle Systems}
(McGraw-Hill, New York, 1971), p.~261.

\bibitem{NO88} John~W.~Negele and Henri~Orland, 
{\it Quantum Many-Particle Systems} (Addison-Wesley, New York, 1988).

\bibitem{Mah90}
G.~D.~Mahan, {\it Many-Particle Physics} (Plenum Press, New York, 1990),
p.~176.

\bibitem{densmat} A.H. MacDonald and S.M. Girvin, 
Phys. Rev. B {\bf 38}, 6295 (1988).

\bibitem{AU74}
T.~Ando, Y.~Uemura, Journ.~Phys.~Soc.~Japan, {\bf 37}, 1044 (1974).

\bibitem{UNHF90}
A.~Usher, R.~J.~Nicholas, J.~J.~Harris, and C.~T.~Foxon, Phys.~Rev.~B {\bf 41}, 
1129 (1990). 

\bibitem{AM91}
D.~Antoniou, A.~H.~MacDonald,  Phys.~ Rev.~B {\bf 43}, 11686 (1991). 
This paper examines the influence of disorder on the dynamic 
susceptibility.  

\bibitem{stoner} E.C. Stoner, Phil. Mag. {\bf 17}, 1018 (1933);
J.C. Slater, Phys. Rev. {\bf 49}, 537 (1936).

\bibitem{korenman} See for example A.T. Alfred, Phys. Rev. {\bf B11},
2597 (1975); G. Lonzarich and A.V. Gold, Can. J. Phys. {\bf 52}, 694
(1974); D.M. Edwards, Can. J. Phys. {\bf 52}, 704 (1974);
V. Korenman, J.L. Murray, and R.E. Prange, Phys. Rev. 
B {\bf 16}, 4032 (1977) and work cited therein.

\bibitem{HE73}
J.~A.~Hertz, D.~M.~Edwards, Journ. Phys. F {\bf 3}, 2174 (1973), 
{\em ibid.} {\bf 3}, 2191 (1973).

\bibitem{SMG92} The influence of dynamic screening 
on exchange-enhanced spin-splitting of Landau levels at 
moderate field strengths has been studied by 
A.~P.~Smith, A.~H.~MacDonald, G.~Gumbs,
Phys.~Rev.~B {\bf 45}, 8829 (1992). 

\bibitem{kunyang} 
Kun Yang and A.H. MacDonald, Phys. Rev. B {\bf 51}, 17247 (1995).

\bibitem{prljmr} Spectral function evolution with temperature does
however play a role in the junction magnetoresistance of magnetic 
tunnel junctions: A.H. MacDonald, T. Jungwirth, and M. Kasner, 
Phys. Rev. Lett.~{\bf 81}, 705 (1998).

\bibitem{CP96}
Similar results have been reported by 
T.~Chakraborty, P.~Pietil\"ainen, 
Phys.~Rev.~Lett.~ {\bf 76}, 4018 (1996); T.~Chakraborty,
P.~Pietil\"ainen, R.~Shankar, 
Europhys. Lett. {\bf 38}, 141 (1997).

\bibitem{TBDPW95}
The given form factor $F(q,w)$ is based on a symmetric charge 
distribution $\rho(z) \sim sin^2(\pi z/w)$ for the electrons 
in a quantum well of width $w$, see: 
R.~Tycko, S.~E.~Barrett, G.~Dabbagh, L.~N.~Pfeiffer, K.~W.~West,
Science {\bf 268}, 1460 (1995).

\bibitem{finitewidth} See for example T. Ando, A.B. Fowler, and F. Stern,
Rev. Mod. Phys. {\bf 54}, 437 (1982) and  A.H. MacDonald and 
G.C. Aers, Phys. Rev. B {\bf 29}, 5976 (1984). 

\bibitem{FBCM94}
H.~A.~Fertig, L.~Brey, R.~C\^ot\'{e}, A.~H.~MacDonald, 
Phys.~Rev.~B {\bf 50}, 11018, (1994). 

\bibitem{slichter} C.P. Slichter, {\it Principles of Magnetic 
Resonance}, (Springer-Verlag, Berlin, 1990).

\bibitem{berg} A. Berg, M. Dobers, R.R. Gerhardts, and K. v. Klitzing,
Phys. Rev. Lett. {\bf 64}, 2563 (1990).

\bibitem{vagner} I.D. Vagner, T. Maniv, Phys. Rev. Lett.
{\bf 61}, 1400 (1988); T. Maniv and I.D. Vagner, Surf. Sci.
{\bf 229}, 134 (1990).

\bibitem{2d2dtunthry} L. Zheng and A.H. MacDonald, Phys.
Rev. B {\bf 47}, 10619 (1993); T. Jungwirth and A.H. MacDonald,
Phys. Rev. B {\bf 53}, 7403 (1996); L. Zheng and S. DasSarma,
Phys. Rev. B {\bf 53}, 9964 (1996).

\bibitem{2d2dtunexpt} J.P. Eisenstein, L.N. Pfeiffer, and K.W. West,
Appl. Phys. Lett. {\bf 58}, 1497 (1991);
J. Smoliner, E. Gornik and G. Weimann, Appl. Phys. Lett.
{\bf 52}, 2136 (1988); 
J.P. Eisenstein, T.J. Gramila, L.N. Pfeiffer and K.W.
West,  Phys. Rev. B {\bf 44}, 6511 (1991); 
R.C. Ashoori and R.H. Silsbee, Solid State Commun.
{\bf 81}, 821 (1992); S.Q. Murphy, J.P.
Eisenstein, L.N. Pfeiffer, and K.W. West,
Phys. Rev. B {\bf 52}, 14825 (1995).

\bibitem{strfieldthry} P. Johansson and J.M. Kinaret, Phys. Rev. Lett.
{\bf 71}, 1435 (1993); A.L. Efros and F.G. Pikus, Phys. Rev. B
{\bf 48}, 14694 (1993); C.M. Varma, A.I. Larkin, and E. Abrahams,
Phys. Rev. B {\bf 49}, 13999 (1994); I.L. Aleiner, H.U. Baranger,
and L.I. Glazman, Phys. Rev. Lett. {\bf 74}, 3435 (1995); 
Rudolf Haussmann, Hiroyuki Mori, and A.H. MacDonald,
Phys. Rev. Lett. {\bf 76}, 979 (1996).

\bibitem{strfieldexpt} J.P. Eisenstein, L.N. Pfeiffer,
and K.W. West, Phys. Rev. Lett. {\bf 74}, 1419 (1995);
R.C. Ashoori, J.A. Lebens, N.P. Bigelow,
and R.H. Silsbee, Phys. Rev. B {\bf 48}, 4616 (1993);
K.M. Brown, N. Turner, J.T. Nicholls, E.H. Linfield,
M. Pepper, D.A. Ritchie, and G.A.C. Jones, Phys. Rev. B {\bf 50}, 15465
(1994). 

\bibitem{Bay62}
G.~Baym, Phys.~Rev.~ {\bf 127}, 1391 (1962); 
G.~Baym, L.~P.~Kadanoff, Phys.~Rev.~ {\bf 124}, 287 (1961). 

\bibitem{const}
We choose for the numerical evaluation of our expressions the value 
$V_0= \sqrt{\pi}/2$, which equals the zero momentum pseudopotential 
in the expansion of the Coulomb interaction in the LLL. In contrast 
to the Coulomb model there is no neutralizing background in the 
hard core interaction model. Therefore 
tadpole diagrams have to be considered. 

\bibitem{Kas97}
M.~Kasner, Physica E {\bf 1}, 71 (1997). 

\bibitem{finsiz}
For a noninteracting system this finite size correction can be exactly 
calculated. We find the value $1/(2N-1), N$ - particle number, 
see the inset in Fig.~20.

\bibitem{AB91}
Recent applications of diagrammatic methods to electrons in 
a strong magnetic field can be found, {\it e.g}, in 
A.~V.~Andreev, Yu.~A.~Bychkov, \v{Z}.~Eksp.~Teor.~Fiz.\ {\bf 100} 752, (1991), 
(Sov.~Phys.~JETP {\bf 73} 404, (1991)); Lian Zheng, A.~H.~MacDonald,
Surface Science {\bf 305} 101, (1994).

\end{thebibliography}
\end{document}